\documentclass{sig-alternate} 

\usepackage{amssymb}
\usepackage[linesnumbered, vlined, ruled]{algorithm2e}
\usepackage{color}
\usepackage{booktabs}
\usepackage{url}
\usepackage{multirow}
\usepackage{textcomp}

\usepackage{amsthm}

\usepackage{enumitem}

\newcommand{\eat}[1]{}
\newcommand{\alt}[2]{#2}
\newcommand{\stitle}[1]{\vspace{0.05cm}\noindent\textbf{#1}}

\newcommand{\la}[1]{\ensuremath{\lambda(#1)}}
\newcommand{\lae}{\ensuremath{\lambda}}
\newcommand{\<}{\ensuremath{\langle}}
\renewcommand{\>}{\ensuremath{\rangle}}

\newcommand{\E}{\ensuremath{\mathcal{E}}}

\newcounter{prob}\newenvironment{myprob}[1][]
{\refstepcounter{prob}\par\setlength{\leftskip}{3pt}\setlength{\rightskip}{3pt}\smallskip\noindent\ignorespaces
   \textbf{Problem~\theprob. [#1]}}
{\smallskip\par}

\newtheorem{definition}{Definition}
\newtheorem{theorem}{Theorem}
\newtheorem{lemma}{Lemma}

\newtheoremstyle{mystyle}% name
  {3pt}%      Space above
  {3pt}%      Space below
  {}%         Body font
  {}%         Indent amount (empty = no indent, \parindent = para indent)
  {\bfseries}% Thm head font
  {.}%        Punctuation after thm head
  {.5em}%     Space after thm head: " " = normal interword space;
  {}%         Thm head spec (can be left empty, meaning `normal')

\theoremstyle{mystyle}
\newtheorem{example}{Example}

\begin{document}

\toappear{

\alt{
Permission to make digital or hard copies of all or part of this work for personal or classroom use is granted without fee provided that copies are not made or distributed for profit or commercial advantage and that copies bear this notice and the full citation on the first page. Copyrights for components of this work owned by others than the author(s) must be honored. Abstracting with credit is permitted. To copy otherwise, or republish, to post on servers or to redistribute to lists, requires prior specific permission and/or a fee. Request permissions from Permissions@acm.org.

SIGSPATIAL'13, November 05 - 08 2013, Orlando, FL, USA\\
Copyright is held by the owner/author(s). Publication rights licensed to ACM.\\
ACM 978-1-4503-2521-9/13/11\$15.00.\\
http://dx.doi.org/10.1145/2525314.2525362
}{}
}

\title{Routing Directions: Keeping it Fast and Simple}

\numberofauthors{2} 

\author{
\alignauthor Dimitris Sacharidis\\
\affaddr{Institute for the Mgmt. of Information Systems}\\
\affaddr{``Athena'' Research Center}\\
\affaddr{Athens, Greece}\\
\email{dsachar@imis.athena-innovation.gr}
\alignauthor Panagiotis Bouros\\
\affaddr{Department of Computer Science}\\
\affaddr{Humboldt-Universit\"{a}t zu Berlin}\\
\affaddr{Berlin, Germany}\\
\email{bourospa@informatik.hu-berlin.de}
}

\setlength\floatsep{1.25\baselineskip plus 3pt minus 2pt}
\setlength\textfloatsep{1.25\baselineskip plus 3pt minus 2pt}
\setlength\intextsep{1.25\baselineskip plus 3pt minus 2 pt}

\maketitle

\begin{abstract}

The problem of providing meaningful routing directions over road networks is
of great importance. In many real-life cases, the fastest route may not be the
ideal choice for providing directions in written/spoken text, or for an
unfamiliar neighborhood, or in cases of emergency. Rather, it is often more
preferable to offer ``simple'' directions that are easy to memorize, explain,
understand or follow. However, there exist cases where the simplest route is
considerably longer than the fastest. This paper tries to address this issue,
by finding near-simplest routes which are as short as possible and 
near-fastest routes which are as simple as possible. Particularly, we focus on
efficiency, and propose novel algorithms, which are theoretically and
experimentally shown to be significantly faster than existing approaches.

\end{abstract}




%!TEX root = main.tex

\section{Introduction}

Finding the fastest route on road networks has received a renewed interest in
the recent past, thanks in large part to the proliferation of mobile 
location-aware devices.  However, there exist many real-life scenarios in which the
fastest route may not be the ideal choice when providing routing directions.

\alt{
\newpage
}
{}

As a motivating example, consider the case of a tourist asking for driving
directions to a specific landmark. Since the tourist may not be familiar with
the neighborhood, it makes more sense to offer directions that involve as few
turns as possible, instead of describing in detail an elaborate fastest route.
As another example, consider an emergency situation, e.g., natural disaster,
terrorist attack, which requires an evacuation plan to be communicated to
people on the site. Under such circumstances of distress and disorganization,
it is often desirable to provide concise, easy to memorize, and clear to
follow instructions.

In both scenarios, the \emph{simplest route} may be more preferable than the
fastest route. As per the most common interpretation \cite{W02}, turns (road
changes) are assigned costs, and the simplest route is the one that has the
lowest total turn cost, termed \emph{complexity}. For simplicity, in the
remainder of this work, we assume that all turns have equal cost equal to 1;
the generalization to non-uniform costs is straightforward.

In some road networks, the simplest and the fastest route may be two
completely different routes. Consider for example a large city, e.g., Paris,
that has a large ring road encircling a dense system of streets. The simplest
route between two nodes that lie on (or are close to) the ring, would be to
follow the ring. On the other hand, the fastest route may involve traveling
completely within the enclosing ring. As a result the length of the simplest
route can be much larger than that of the fastest route, and vice versa.

Surprisingly, with the exception of \cite{JL11}, the trade-off between length
and complexity in finding an optimal route has not received sufficient
attention. Our work addresses this issue by studying the problem of finding
routes that are as fast and as simple as possible.

In particular, we first study the \emph{fastest simplest problem}, i.e., of
finding the fastest among all simplest routes, which was the topic of
\cite{JL11}. We show that although, for this problem, a label-setting method
(a variant of the basic Dijkstra's algorithm) cannot be directly applied on
the road network, it is possible to devise a conceptual graph on which it can.
In fact, our proposed algorithm is orders of magnitude faster than the
baseline solution. Moreover, using a similar methodology, it is possible to
efficiently solve the \emph{simplest fastest problem}.

Subsequently, we investigate the length-complexity trade-off and introduce two
novel problems that relax the constraint that the returned routes must be
either fastest or simplest. The \emph{fastest near-simplest problem} is to
find the fastest possible route whose complexity is not more than $1+\epsilon$
times larger than that of the simplest route. On the other hand, the
\emph{simplest near-fastest problem} is to find the simplest possible route
whose length is not more than $1+\epsilon$ times larger than that of the
fastest route.

These near-optimal problems are significantly more difficult to solve compared
to their optimal counterparts. The reason is that there cannot exist a
principle of optimality, exactly because the requested routes are by
definition \emph{sub- optimal} in length and complexity. Therefore, one must
exhaustively enumerate all routes, and only hope to devise pruning criteria to
quickly discard unpromising sub-routes.

We propose two algorithms, based on route enumeration, for finding the
simplest near-fastest route; their extension for the fastest near-simplest
problem is straightforward. The first follows a depth-first search principle
in enumerating paths, whereas the second is inspired by A$^*$ search. Both
algorithms apply elaborate pruning criteria to eliminate from consideration a
large number of sub-routes. Our experimental study shows that they run in less
than 400 msec in networks of around 80,000 roads and 110,000 intesections.

The remainder of the paper is organized as follows. Section~\ref{sec:prelim}
formally defines the problems and reviews related work. Section~\ref{sec:simplest}
discusses the fastest simplest, and Section~\ref{sec:near-fastest} the
simplest near-fastest problem. Then, Section~\ref{sec:exps} presents our
experimental study and Section~\ref{sec:concl} concludes the paper.

%!TEX root = main.tex

\section{Preliminaries}
\label{sec:prelim}

Section~\ref{sec:definitions} presents the necessary definition, while
Section~\ref{sec:related} reviews relevant literature.

\subsection{Definitions}
\label{sec:definitions}

Let $V$ denote a set of nodes representing road intersections. A \emph{road}
$r$ is a sequence of distinct nodes from $V$. Let $R$ denote a set of roads,
such that all nodes appear in at least one road, and any pair of consecutive
nodes of some road do not appear in any other, i.e., the roads do not have
overlapping subsequences. For a node $n \in V$, the notation $R(n) \subseteq
R$ represents the non-empty subset of roads that contain $n$. For two
consecutive nodes $n_i$, $n_j$ of some road $r$, the notation $R(n_i, n_j)$ is
a shorthand for $r$.

\begin{definition}

The \emph{road network} of $R$ is the directed graph $G_R(V,E)$, where $V$ is
the set of nodes, and $E \subseteq V \! \times \! V$ contains an edge $e_{ij} =
(n_i,n_j)$ if $n_i$, $n_j$ are consecutive nodes in some road.

\end{definition}

A road network is associated with two cost functions. The \emph{length
function} $L$ assigns to each edge a cost representing its length, i.e., the
travel time or distance between them; formally, $L: E \to \mathbb{R}^+$ maps
each edge $(n_i,n_j)$ to the length $L(n_i, n_j)$ of the road segment $n_i$ to
$n_j$.

The \emph{complexity function} $C$ assigns to each turn from road $r_i$ to
road $r_j$ via node $n_x$, which lies on both $r_i$ and $r_j$, the cost of
making the turn. Formally, $C: V\! \times\! R \!\times\! R \to \mathbb{R}^+$ maps $(n_x,r_i,r_j)$ to complexity $C(n_x,r_i,r_j)$ from $r_i$ to $r_j$
via $n_x$.

A \emph{route} $\rho = (n_a, n_b, \dots)$ is a path on graph $G_R$, i.e., a
sequence  of nodes from $V$, such that for any two consecutive nodes, say
$n_i$, $n_j$, there exists an edge $e_{ij}$ in $E$.

The \emph{length} $L(\rho)$ of a route $\rho$ is the sum of the lengths for
each edge it contains, and represents the total travel time or distance
covered along this route; formally,
\begin{equation}
L(\rho) = \sum_{(n_i,n_j) \in \rho} L(n_i, n_j).
\label{eqn:length}
\end{equation}

A route from source $n_s$ to target $n_t$ is called a
\emph{fastest route} if its length is equal to the smallest length of any
route from $n_s$ to $n_t$. Given a parameter $\epsilon$, a route from $n_s$ to
$n_t$ is called an \emph {near-fastest route} if its length is at most
$(1+\epsilon)$ times that of the fastest route from $n_s$ to $n_t$.

The \emph{complexity} $C(\rho)$ of a route $\rho$ is the sum of complexities for
each turn it contains; formally
\begin{equation}
C(\rho) = \sum_{(n_i,n_j,n_k) \in \rho} C(n_j, R(n_i, n_j), R(n_j, n_k)),
\label{eqn:complex}
\end{equation}
where $n_i$, $n_j$, $n_k$ are three consecutive nodes in $\rho$, and $R(n_i,
n_j)$, $R(n_j, n_k)$ are the (unique) roads containing segments $(n_i, n_j)$
and $(n_j, n_k)$, respectively. A route from $n_s$ to $n_t$ is called a
\emph{simplest route} if its complexity is equal to the lowest complexity of
any route from $n_s$ to $n_t$. Given a parameter $\epsilon$, a route from
$n_s$ to $n_t$ is called a \emph {near-simplest route} if its complexity is
at most $(1+\epsilon)$ times that of the simplest route from $n_s$ to $n_t$.
Note that the complexity of a simplest route can be 0, i.e., when no
road changes exist. In this case, all near-simplest routes must also have
complexity 0. To address this, one could simply change the definition of
complexity to be the number of roads in a route, and thus at least 1. In the
remainder of this paper, we ignore this case, and simply use the original 
definition of complexity.

\begin{figure}
\centering
\includegraphics[width=4.2cm]{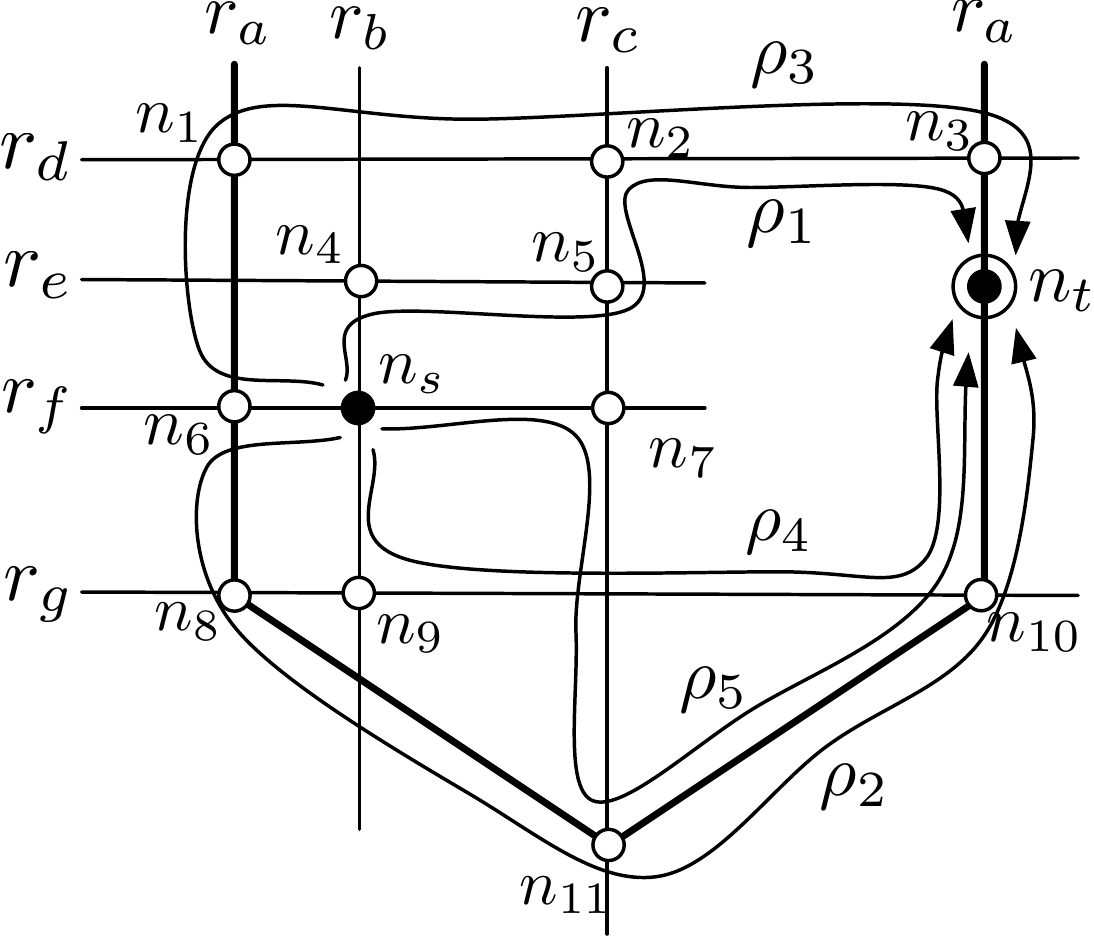}
\vspace{-10pt}
\caption{An example road network of seven roads $r_a$ -- $r_g$, where five routes $\rho_1$ -- $\rho_5$ from node $n_s$ to $n_t$ are depicted.}
\vspace{-5pt}
\label{fig:routes}
\end{figure}

\begin{table}
\scriptsize
\centering
\caption{Costs of routes in Figure~\ref{fig:routes}}
\vspace{5pt}
\label{tab:routes}
\begin{tabular}{cccc}
\toprule
\textbf{road} & \textbf{length} & \textbf{complexity} & \textbf{type}\\
\midrule
$\rho_1$ & 10 & 4 & SF\\
$\rho_2$ & 40 & 1 & FS\\
$\rho_3$ & 20 & 3 & SNF ($\epsilon=1$)\\
$\rho_4$ & 30 & 2 & FNS ($\epsilon=1$)\\
$\rho_5$ & 40 & 2 & ---\\
\bottomrule
\end{tabular}
\end{table}

This work deals with the following problems. To the best of our knowledge only
the first has been studied before in literature \cite{JL11}.

\begin{myprob}[Fastest Simplest Route]
Given a source $n_s$ and a target $n_t$, find a route that has the smallest
length among all simplest routes from $n_s$ to $n_t$.
\label{prb:fs}
\end{myprob}

\begin{myprob}[Simplest Fastest Route]
Given a source $n_s$ and a target $n_t$, find a route that has the smallest
complexity among all fastest routes from $n_s$ to $n_t$.
\label{prb:sf}
\end{myprob}

\begin{myprob}[Fastest Near-Simplest Route]
Given a source $n_s$ and a target $n_t$, find a route that has the smallest
length among all near-simplest routes from $n_s$ to $n_t$.
\label{prb:fns}
\end{myprob}

\begin{myprob}[Simplest Near-Fastest Route]
Given a source $n_s$ and a target $n_t$, find a route that has the lowest
complexity among all near-fastest routes from $n_s$ to $n_t$.
\label{prb:snf}
\end{myprob}

Note that the first two problems are equivalent to the last two, respectively,
if we set $\epsilon = 0$. We next present an example illustrating these
problems.

\begin{example}

Consider the road network of
Figure~\ref{fig:routes} consisting of 7 two-way roads $r_a$ -- $r_g$. Note
that all roads have either a north-south or an east-west direction, except
road $r_a$, which is a ring-road and is thus depicted with a stronger line.
The figure also portrays 11 road intersections $n_1$--$n_{11}$ with hollow
circles, and two special nodes, the source $n_s$, drawn with filled circle, and
the target $n_t$, drawn with a filled circle inside a larger hollow one.

Next, consider five possible routes $\rho_1$--$\rho_5$ starting from $n_s$ and
ending at $n_t$, which are drawn in Figure~\ref{fig:routes}, and whose lengths
and complexities are shown in Table~\ref{tab:routes}. Observe that $\rho_1$ is
the fastest route from $n_s$ to $n_t$ with length 10, and, moreover, it has
the lowest complexity 4 among all other fastest route (no other exists). Thus,
$\rho_1$ is the simplest fastest route and the answer to Problem~\ref{prb:sf}.

On the other hand, $\rho_2$ is the fastest simplest route and the answer to
Problem~\ref{prb:fs}, as it has the lowest complexity 1, following the ring-road to reach the target. But its length, 40, is quite large
compared to the other possible routes.

Assume $\epsilon = 1$ for the complexity, so that a near-simplest route can
have complexity at most twice that of the simplest route, i.e., 2. Observe
that two routes $\rho_4$ and $\rho_5$ are near-simplest. Among them $\rho$ is
the fastest, and is thus the answer to Problem~\ref{prb:fns}.

Moreover, assume $\epsilon = 1$ for the length as well, so that a near-fastest
route can have length at most twice that of the fastest route, i.e., 20.
Notice that only $\rho_3$ is near-fastest and thus is the answer to
Problem~\ref{prb:snf}.

Observe that if we set $\epsilon = 2$ for the length, near-fastest routes can
have length as large as 30. In this case, $\rho_3$ and $\rho_4$ are near-
fastest, with the latter being the simplest near-fastest.
\qed
\end{example}

\subsection{Related Work}
\label{sec:related}

Dijkstra \cite{D59} showed that the fastest route problem exhibits a principle
of sub-route optimality and proposed its famous dynamic programming method for
finding all fastest routes from a given source. Bi-directional search
\cite{N66}, i.e., initiating two parallel searches from the source and the
target can significantly expedite finding the fastest source-to-target route.
Since this early work around the 60's, numerous network preprocessing
techniques exist today, including landmarks \cite{GH05}, reach \cite{G04},
multi-level graphs \cite{SWZ02}, graph hierarchies \cite{SS05, GSSD08}, graph
partitioning \cite{MSS+07}, labelings \cite{ADGW11}, and their combinations
\cite{BD09,BDS+10,DGNW12}, which are capable of speeding up Dijkstra's
algorithm by orders of magnitude in several instances.

The problem of finding the simplest route was first studied in \cite{C61}, and
more recently in \cite{W02,DK03}. The basic idea behind these methods, is to
construct a pseudo-dual graph of the road network, where road segments become
the nodes, the turns between two consecutive road segments become the edges,
which are assigned turn costs. Then, finding the simplest route reduces to
finding the shortest path on the transformed graph. In contrast, the recent
work of \cite{GV11} solves the simplest route problem directly on the road
network. Note that Problem~\ref{prb:fs} differs with respect to the simplest
route problem, as it request a specific simplest route, that with the smallest
length. The aforementioned methods return any simplest route.

To the best of our knowledge, only the work in \cite{JL11} addresses
Problem~\ref{prb:fs}, as it proposes a solution that first finds all simplest
routes and then selects the fastest among them. The proposed method serves as
a baseline approach to our solution for Problem~\ref{prb:fs} and is detailed
in Section~\ref{sec:baseline}.

Problems~\ref{prb:fns} and \ref{prb:snf} are related to the problem of finding
the near-shortest paths on graphs \cite{BW84,MW05}. This paper differs with
that line of work in two ways. First, the studied problems involve two cost
metrics, length and complexity. Second, their solution is a single route,
instead of all possible near-optimal routes.

Problems~\ref{prb:fns} and \ref{prb:snf} are also related to multi-objective
shortest path problems (see e.g., \cite{RE09,KRS10}), which specify more than
one criteria and may return sub-optimal routes. This paper differs with that
line of work again in two ways. First, they request a single route. Second,
the studied problems introduce a hard constraint on the length or complexity
of a solution. Nonetheless, an interesting extension to our work would be to
return all routes that capture different length-complexity trade-offs.

%!TEX root = main.tex

\section{Fastest Simplest Route}
\label{sec:simplest}

This section discusses Problem~\ref{prb:fs}, and introduces an algorithm that
takes advantage of the principle of sub-route optimality to expedite the search.
Solving Problem~\ref{prb:sf} is similar, and thus details are omitted. We
first present a recent baseline solution in Section~\ref{sec:baseline}, and
then discuss our approach in Section~\ref{sec:FS}.

\subsection{Baseline Solution}
\label{sec:baseline}

The work in \cite{JL11} was the first to address the fact that there can exist
multiple simplest routes with greatly varying length, and proposes a solution
to finding the fastest among them. This method, which we denote as BSL,
operates on a graph that models the \emph{intersections} of the roads in $R$.

\begin{definition}

The \emph{intersection graph} of $R$ is the undirected graph $G_I(R,I)$, where
$R$ is the set of roads; and $I \subseteq R \times R \times V$ contains
intersection $(n_x,r_i, r_j)$ if $r_i \in R(n_x)$ and $r_j \in R(n_x)$, i.e.,
node $n_x$ belongs to both roads $r_i$ and $r_j$.

\end{definition}

A path on the intersection graph, i.e., a sequence of $R$ vertices such that
there exists an intersection in $I$ for any two consecutive vertices in the
sequence, is called a \emph{road sequence}.

BSL finds the fastest simplest route from a source node $n_s$ to a target node
$n_t$. It is based on the observation that a simplest route from $n_s$ to
$n_t$ in the road network is related to a shortest road sequence from a road
that contains $n_s$ to one that contains $n_t$ in the intersection graph. More
precisely, BSL operates as follows.

\begin{enumerate}[leftmargin=12pt]
\itemsep0em

\item For each source road in $R(n_s)$, find the number of intersections of
the shortest road sequence from that source to any of the target roads in
$R(n_t)$, e.g., using a single-source shortest path algorithm on the
intersection graph.

\item Determine the smallest number of intersections among those found in the
previous step. This number corresponds to the fewest possible intersections in
a road sequence that starts from a source and ends at a target road, and is
thus equal to the complexity of the simplest route from $n_s$ to $n_t$ plus 1.

\item Enumerate (e.g., using depth-limited dfs) all road sequences from a
source to a target road that have exactly as many intersections as the number
determined in the previous step. For each road sequence produced, convert it
to a route and determine its length.

\item Select the route with the minimum length, i.e., the fastest, among those
produced in the previous step.
 
\end{enumerate}

\subsection{The FastestSimplest Algorithm}
\label{sec:FS}

The proposed algorithm operates directly on the road network. However, a
direct application of a Dijkstra-like (label-setting \cite{D59}) method is not possible,
because the principle of sub-route optimality does not hold. In particular,
this principle suggests that if node $n_x$ is in the fastest simplest route
from $n_s$ to $n_t$ then \emph{any} fastest simplest sub-route from $n_s$ to
$n_x$ can be extended to a fastest simplest route from $n_x$ to $n_t$. In
comparison, it is easy to see that the principle holds for fastest route
(shortest paths), as \emph{all} fastest sub-routes can be extended to fastest
routes.

We give a counter-example for the principle of optimality on fastest simplest
routes using the road network of Figure~\ref{fig:routes}. Consider the routes $\rho_2 \! = \! (n_s, n_6, n_8, n_{11}, n_{10}, n_t)$ and $\rho_5 =
(n_s, n_7, n_{11}, n_{10}, n_t)$. The sub-route $\rho_5' = (n_s, n_7, n_{11})$
of $\rho_5$ has length 20 and complexity 1, as it involves a single turn from
road $r_f$ to $r_c$ via node $n_7$. Similarly, the sub-route $\rho_2' = (n_s,
n_6, n_8, n_{11})$ of $\rho_2$ has length 20 and complexity 1, as it involves
a single turn from $r_f$ to $r_a$ via node $n_6$. Therefore, both sub-routes
are fastest simplest from $n_s$ to $n_{11}$. However, the extension of
$\rho_5'$ does not give a fastest simplest route from $n_s$ to $n_t$; $\rho_5$
makes an additional turn at node $n_{11}$ compared to $\rho_2$. This violates
the principle of optimality for fastest simplest routes.

Therefore, a Dijkstra-like method, which directly exploits this
principle of optimality, cannot be applied. For instance, such a method could
reach $n_{11}$ first via $r_c$ and subsequently ignore any other path reaching
$n_{11}$, including the sub-route via $r_a$, and thus missing the optimal route
from $n_s$ to $n_t$.

To address the aforementioned lack of sub-route optimality, we construct a
conceptual expanded graph, on which the principle optimality holds.
Additionally, we show that expanded routes on this graph are uniquely
associated with routes on the road network. We emphasize that the expanded
graph is only a conceptual structure used for presentation purposes, and that
the proposed algorithm does not make use of it as it operates directly on the
road network.

\begin{definition}

The \emph{expanded graph} of $G_R(V,E)$ is the directed graph
$G_\E(V',E')$, where $V' \subseteq V \times R$ contains an expanded node $(n_x, r_i)$
if $r_i \in R(n_x)$; $E' \subseteq V' \times V'$ contains an edge $( (n_x,r_i)
, (n_y,r_j))$ if $r_i \in R(n_x)$ and $r_j \in R(n_x)$ ($r_i$ and $r_j$ could
be the same road), and additionally $n_x$, $n_y$ are consecutive nodes in
$r_j$.

\end{definition}

An expanded route $\rho_\E = ((n_a,r_i) , (n_b,r_j), \dots)$ is a path on the
expanded graph $G_\E$. Each expanded edge can be associated with a
length and a turn cost. Therefore, it is possible to define the following
costs for an expanded route.

The length $L(\rho_\E)$ of an expanded route $\rho_\E$ is the sum of the
lengths associated with each expanded edge; formally,
\begin{equation}
L(\rho_\E) = \sum_{((n_x,r_i), (n_y,r_j)) \in \rho_\E} L(n_x, n_y).
\label{eqn:length_exp}
\end{equation}

Similarly, the complexity $C(\rho_\E)$ of an expanded route $\rho_\E$ is the
sum of the turn costs associated with each expanded edge; formally,
\begin{equation}
C(\rho_\E) = \sum_{((n_x,r_i), (n_y,r_j)) \in \rho_\E} C(n_x, r_i, r_j).
\label{eqn:complex_exp}
\end{equation}

An important property regarding the length and complexity of an expanded route is the following. Note that a similar propery does not generally hold for routes on the road network $G_R$.

\begin{lemma}

Let $\rho_{\E}^1$ be an expanded route from $(n_s, r_i)$ to $(n_x, r_y)$, and
$\rho_{\E}^2$ be an expanded route from $(n_x, r_y)$ to $(n_t, r_j)$. If
$\rho_{\E}^1\rho_{\E}^2$ denotes the concatenation of the two expanded routes,
then, it holds that $L(\rho_{\E}^1\rho_{\E}^2) = L(\rho_{\E}^1)+
L(\rho_{\E}^2)$, and $C(\rho_{\E}^1\rho_{\E}^2) = C(\rho_{\E}^1)+
C(\rho_{\E}^2)$.

\label{lem:distribute}
\end{lemma}

\alt{
Please note that the proofs of all lemmas and theorems can be found in the
technical report \cite{SB13}.
}
{
\begin{proof}
The proof follows because both the length and complexity functions are defined
independently for each expanded edge of an expanded route.
\end{proof}
}

It should be apparent that expanded routes are closely related with (non-expanded) routes.
First, let us examine the $G_R$ to $G_\E$ relationship, which is one to many.

We associate a route $\rho$ from $n_s$ to $n_t$ on the road network $G_R$ to a
set $\E(\rho)$ of expanded routes on $G_\E$, which only differ in their first
and last expanded nodes. Particularly, for an expanded route $\rho_\E \in
\E(\rho)$, its first expanded node is $(n_s, r_i)$, where $r_i \in R(n_s)$,
the last expanded node is $(n_t, r_j)$, where $r_j \in R(n_t)$, and the $k$-th
expanded node (for $k>1$) is $(n_k, R(n_{k-1}, n_k))$, where $n_{k-1}$, $n_k$
are the $(k\!-\!1)$-th, $k$-th nodes in $\rho$, respectively, and $R(n_{k-1},
n_k)$ is the unique road that contains the edge $(n_{k-1}, n_k)$. Conversely,
an expanded route $\rho_\E$ is associated with a unique route $\rho$.

Given a route $\rho$, we define the \emph{special expanded route} of $\rho$,
denoted as $\rho_\E^*$, to be the expanded route in $\E(\rho)$ that has $(n_s,
R(n_s,n_{s+1}))$ as its first expanded node, and $(n_t, R(n_{t-1},n_t))$ as
its last expanded node, where $n_{s+1}$ is the second node in route $\rho$,
$R(n_s,n_{s+1})$ is the unique road containing edge $(n_s,n_{s+1})$, $n_{t-1}$
is the second-to-last node in route $\rho$, and $R(n_{t-1},n_t)$ is the unique
road containing edge $(n_{t-1},n_t)$.

An even more important property is the following.

\begin{lemma}

The length of a route $\rho$ is equal to the length of any expanded route
$\rho_\E \in \E(\rho)$. The complexity of a route $\rho$ is equal to the
complexity of the special expanded route $\rho_\E^* \in \E(\rho)$.
\label{lem:expand}
\end{lemma}

\alt{}{
\begin{proof}

For convenience, assume that $\rho = (n_1, n_2, \dots, n_t)$. Then, the length
of the route is $\displaystyle L(\rho) = \sum_{k=1}^{t-1}L(n_k, n_{k+1})$, and
its complexity is $\displaystyle C(\rho) = \sum_{k=2}^{t-1} C(n_k, R(n_{k-1},
n_k), R(n_k, n_{k+1}))$.

An expanded route of $\rho$ is
$$\rho_\E = ( (n_1,r_s), \dots, (n_k,R(n_{k-1},n_k)), \dots, (n_t, r_e) ),$$
where $r_s \in R(n_1)$ and $r_e \in R(n_t)$. Observe that the length of an
expanded route is $L(\rho_\E) = \sum_{k=1}^{t-1} L(n_k, n_{k+1})$, which is
equal to $L(\rho)$. Hence, the first part of the lemma holds.

The special expanded route of $\rho$ is $\rho_\E^* =$
$$( (n_1,R(n_1,n_2)), \dots, (n_k,R(n_{k-1},n_k)), \dots, (n_t, R(n_{t-1},n_t)) ).$$

The complexity of the special expanded route is
\begin{align*}
C(\rho_\E^*) = & \ C(n_1, R(n_1,n_2), R(n_1,n_2)) \\
& \ + \sum_{k=2}^{t-1} C(n_k, R(n_{k-1}, n_k), R(n_k, n_{k+1})) \\
= & \ 0 + C(\rho),
\end{align*}
where the first term is zero because the turn cost on the same road is zero. Hence the second part of the lemma also holds.
\end{proof}
}

Next, let us examine the $G_\E$ to $G_R$ relationship, which is many to one.
We associate an expanded route $\rho_\E$ from $(n_s, r_i)$ to $(n_t, r_j)$ to
a unique route $\rho = \E^{-1}(\rho_\E)$ from $n_s$ to $n_t$ on $G_R$, such
that the $k$-th node (for any $k$) of $\rho$ is $n_k$, where $(n_k, r_x)$ is
the $k$-th expanded node of $\rho_\E$.

\begin{lemma}

The length of an expanded route $\rho_\E$ is equal to the length of the route
$\rho = \E^{-1}(\rho_\E)$. The complexity of a route $\rho_\E$ is not smaller
than the complexity of the route $\rho = \E^{-1}(\rho_\E)$.

\label{lem:contract}
\end{lemma}

\alt{}{
\begin{proof}

For convenience, assume that the expanded route is $\rho_\E = ( (n_1, r_1),
(n_2, r_2), \dots, (n_t, r_t))$. Then, its length is $\displaystyle L(\rho_\E)
= \sum_{k=1}^{t-1}L(n_k, n_{k+1})$, and its complexity is $\displaystyle
C(\rho_\E) = \sum_{k=1}^{t-1} C(n_k, r_k, r_{k+1})$.

The route on $G_R$ is
$\rho = \E^{-1}(\rho_\E) = (n_1, n_2, \dots, n_t)$, and has length
$\displaystyle L(\rho) = \sum_{k=1}^{t-1}L(n_k, n_{k+1}) = L(\rho_\E)$, which proves the first part of the lemma.

The complexity of the non expanded route $\rho$ is 
\begin{align*}
C(\rho) = & \ \sum_{k=2}^{t-1} C(n_k, R(n_{k-1}, n_k), R(n_k, n_{k+1}))\\
= & \sum_{k=2}^{t-1} C(n_k, r_k, r_{k+1})\\
= & \sum_{k=1}^{t-1} C(n_k, r_k, r_{k+1}) - C(n_1, r_1, r_2)\\
= & C(\rho_\E) - C(n_1, r_1, r_2) \leq C(\rho_\E),
\end{align*}
since $R(n_{k-1}, n_k) =r_k$ and $R(n_k, n_{k+1}) = r_{k+1}$, which proves the second part of the lemma.
\end{proof}
}

We next introduce a lexicographic total order, which applies to routes or
expanded routes. Note that in this section, we use this order exclusively for
expanded routes. Given two routes $\rho^1$, $\rho^2$, we say that $\rho^1$ is
FS-shorter than $\rho^2$ and denote as $\rho^1 <_{FS} \rho^2$ if $C(\rho^1) <
C(\rho^1)$ or if $C(\rho^1) = C(\rho^1)$ and $L(\rho^1) < L(\rho^1)$.
Intuitively, being FS-shorter implies being simpler or as simple but faster.

The following theorem presents an important property regarding this order on expanded routes.

\begin{theorem}

Let $\rho^{FS}$ be a fastest simplest route on $G_R$ from $n_s$ to $n_t$, and
let $\rho_\E^{FS*}$ denote its special expanded route. It holds that there
exists no other expanded route that starts from $(n_s, r_i)$ and ends at
$(n_t, r_j)$, for any $r_i \in R(n_s)$ and $r_j \in R(n_t)$ that is FS-shorter
than $\rho_\E^{FS*}$.
\label{thm:equivalent}

\end{theorem}

\alt{}{
\begin{proof}

We prove by contradiction. Suppose there exists an expanded route $\rho_\E'$
from $(n_s, r_i)$ to $(n_t, r_j)$, for some $r_i \in R(n_s)$ and $r_j \in
R(n_t)$, such that it is FS-shorter than $\rho_\E^{FS*}$. Therefore, one of
the two conditions are true: 
\begin{equation}
C(\rho_\E') < C(\rho_\E^{FS*}), \text{ or}
\label{eqn:cnd1}
\end{equation}
\begin{equation}
C(\rho_\E') = C(\rho_\E^{FS*}) \text{ and } L(\rho_\E') < L(\rho_\E^{FS*}).
\label{eqn:cnd2}
\end{equation}

Consider the route $\rho' = \E^{-1}(\rho_\E')$. From Lemma~\ref{lem:contract},
we have that $L(\rho') = L(\rho_\E')$ and $C(\rho') \leq C(\rho_\E')$.
Moreover, since $\rho_\E^{FS*}$ is the special expanded route of $\rho^{FS}$,
we have from Lemma~\ref{lem:expand} that $L(\rho_\E^{FS*}) = L(\rho^{FS})$ and
$C(\rho_\E^{FS*}) = C(\rho^{FS})$.

Using these relationships, the two conditions become
\begin{equation}
C(\rho') < C(\rho^{FS}), \text{ or}
\end{equation}
\begin{equation}
C(\rho') = C(\rho^{FS}) \text{ and } L(\rho') < L(\rho^{FS}),
\end{equation}
which imply that $\rho'$ is either simpler than $\rho^{FS}$ or as simple but
faster. Therefore, $\rho^{FS}$ cannot be a fastest simplest route, which is a
contradiction.
\end{proof}
}

Theorem~\ref{thm:equivalent} implies that to find a fastest simplest route on $G_R$, it
suffices to find a FS-shortest expanded route on $G_\E$.

The following theorem shows that a principle of optimality holds for
FS-shortest expanded routes on $G_\E$.

\begin{theorem}

Let $\rho_{\E}$ denote an FS-shortest expanded route from $(n_s, r_i)$ to
$(n_t, r_j)$ that passes through $(n_x, r_y)$. Furthermore, let $\rho_{\E}^1$
denote its sub-route from $(n_s, r_i)$ to $(n_x, r_y)$, and $\rho_{\E}^2$ its
sub-route from $(n_x, r_y)$ to $(n_t, r_j)$. It holds that both $\rho_{\E}^1$
and $\rho_{\E}^2$ are FS-shortest. Moreover, if $\rho_{\E}^{1'}$ is another
FS-shortest expanded route from $(n_s, r_i)$ to $(n_x, r_y)$, then
$\rho_{\E}^{1'}\rho_{\E}^2$ is an FS-shortest expanded route from $(n_s, r_i)$
to $(n_t, r_j)$.

\label{thm:optimality}
\end{theorem}

\alt{}{
\begin{proof}

Suppose $\rho_{\E}^1$ is not an FS-shortest expanded route from $(n_s, r_i)$
to $(n_x, r_y)$. Then there exists another expanded route, say
$\rho_{\E}^{1*}$, that is FS-shorter. From Lemma~\ref{lem:distribute}, it is
easy to see that $\rho_{\E}^{1*} \rho_{\E}^2$ is FS-shorter than $\rho_{\E}$,
which is a contradiction as the latter is FS-shortest. A similar argument
holds for $\rho_{\E}^2$. Hence the first part of the theorem is proved.

Regarding expanded route $\rho_{\E}^{1'}$, observe that since it is FS-
shortest it has the same length and complexity with $\rho_{\E}^1$. Then by
Lemma~\ref{lem:distribute}, $\rho_{\E}^{1'}\rho_{\E}^2$  has the same length
and complexity with $\rho_{\E}$. Therefore, it has to be also FS-shortest,
which proves the second part of the theorem.
\end{proof}
}

The key point in Theorem~\ref{thm:optimality} is that it holds for \emph{any}
expanded route on $G_\E$. In contrast, this does not hold for routes on the
road network $G_R$, as we have argued in the beginning of this section.

Given Theorem~\ref{thm:optimality}, we can apply a Dijkstra's algorithm, or
any variant, to find the FS-shortest expanded route on $G_\E$. Then, from
Theorem~\ref{thm:equivalent}, we immediately obtain a fastest simplest route
on $G_R$.

In what follows, we present the FastestSimplest (FS) algorithm, a label-
setting method (a generalization of Dijkstra's algorithm) for finding a
fastest simplest route on $G_R$, which operates directly on the road network
and constructs directly routes, instead of expanded routes.

The pseudocode of the FS algorithm is depicted in
Algorithm~\ref{algo:fs}. Although FS operates on the road network
$G_R$, it updates labels for expanded nodes. A label $\la{n,r}$  for expanded
node $(n,r)$ is equal to $\< n, r | len, cpl, n_{prev}, r_{prev} \>$ and
represents an expanded route from $(n_s, r_i)$, for some $r_i \in R(n_s)$, up
to $(n,r)$. In particular, $len$, $cpl$ are the length and complexity of this
expanded route, while $(n_{prev}, r_{prev})$ is the second-to-last expanded
node. Note that this expanded route is FS-shortest only when it is explicitly
marked as \texttt{final}.

FS uses a minheap $H$ to guide the search, visiting nodes of $G_R$. An entry
of $H$ is a label, and its key is the label's length, complexity pair $(len,
cpl)$. Labels in $H$ are ordered using the FS-shorter total order. At each
iteration, FS deheaps a label, marks it \texttt{final} and advances the
search frontier.

For any road $r_i \in R(n_s)$, the algorithm initializes the heap with the
label $(n_s, r_i | 0, 0, n_{\varnothing}, r_i)$ (lines 1--3). The dummy node
$n_{\varnothing}$ signifies that $n_s$ is the first node in any route
constructed.

The algorithm proceeds iteratively, deheaping labels until the heap is
depleted (line 4), or the label involving the target is deheaped (line 7).
Assume $(n_x, r_i | len, cpl, n_w, r_h)$ is the deheaped label (line 5). As
explained before, this label is finalized (line 6).

If the label does not involve the target, FS expands the current route
(represented by the deheaped label) considering each outgoing edge $(n_x,
n_y)$ of $n_x$ (line 8), and each road $r_j$ that contains $n_y$ (line 9).

If the label $\la{n_y, r_j}$ does not exist (line 10), its label is
initialized with length equal to $len$ plus the distance $L(n_x, n_y)$ of the
outgoing edge, and with complexity equal to $cpl$ plus the complexity $C(n_y,
r_i, r_j)$ of transitioning from road $r_i$ to $r_j$ via node $n_y$ (lines 11
--12).

Otherwise, if label $\la{n_y, r_j}$ exists but is not \texttt{final} (line
13), it is retrieved (line 14). The label will be updated if the extension of
the current expanded route is FS-shorter that the one currently represented in
the label (lines 15--17).

The fastest simplest route can be retrieved with standard backtracking. We
keep all deheaped labels, and then starting from the label containing the
target, we identify the previous expanded node (from the information stored in
the label) and retrieve its label, until the source is reached.

\begin{theorem}

The FS algorithm correctly finds a fastest simplest route from $n_s$ to $n_t$.

\end{theorem}

\alt{}{
\begin{proof}

We first show that FS finds a FS-shortest expanded route, say $\rho_\E^{FS}$,
among those from any $(n_s, r_i)$ to any $(n_t, r_j)$, where $r_i \in R(n_s)$
and $r_j \in R(n_t)$. Consider a virtual expanded node $(n_s, r_\varnothing)$
that has outgoing edges to all $(n_s, r_i)$ for $r_i \in R(n_s)$, with length
and complexity set to 0. Observe that the FS algorithm uses a label-setting
method (Theorem~\ref{thm:optimality}) to find an FS-shortest expanded route
from $(n_s, r_\varnothing)$ to any expanded target node $(n_t, r_j)$, where
$r_j \in R(n_t)$. This expanded route has length and complexity exactly equal
to $\rho_\E^{FS}$.

By Theorem~\ref{thm:equivalent} $\rho_\E^{FS}$ is the special expanded route
of a fastest simplest route from $n_s$ to $n_t$, which concludes the proof.
\end{proof}
}

\stitle{Analysis.} Let $\delta = \max_{n \in V} |R(n)|$ denote the maximum
degree of the road network $G_R$, i.e., the maximum number of roads a node can
belong to. Note that there exist not more than $\delta |V|$ labels, i.e.,
$(n,r)$ pairs. In the worst case, FastestSimplest performs an enheap and
deheap operation for each label. Furthermore, in the worst case,
FastestSimplest examines each edge $\delta$ times, one for each label of a
node. For each examination, it may update $\delta$ labels, in the worst case.
Therefore, there is a total of $\delta^2 |E|$ updates, in the worst case.
Assuming a Fibonacci heap, the time complexity of FastestSimplest is
$O(\delta^2 |E| + \delta |V|\log|V|)$ amortized. Moreover, since the heap may
contain an entry for each label, the space complexity is $O(\delta |V|)$.

\stitle{Discussion.} Thanks to Theorem~\ref{thm:optimality}, the FS algorithm
essentially solves a shortest path problem defined on the expanded graph
directly on the road network. It is thus possible to substitute the underlying
basic label-setting method method with a more efficient variant. Bi-
directional search and  all graph preprocessing techniques, discussed in
Section~\ref{sec:related}, are compatible and can expedite the underlying
method.

\begin{algorithm}[t]
\scriptsize
\SetInd{0.3em}{0.8em}
\SetCommentSty{textsf}
\SetArgSty{textrm}
\DontPrintSemicolon
\SetKw{Break}{break}
\SetKw{Enheap}{enheap}
\SetKw{Deheap}{deheap}
\SetKw{Retrieve}{retrieve}
\SetKw{Mark}{mark}
\SetKw{Update}{update}
\KwIn{road network $G_R$; function $L$; function $C$; source $n_s$; target $n_t$}
\KwOut{length \textit{fsL} and complexity \textit{fsC} of fastest simplest route from $n_s$ to $n_t$}
\SetKwInput{KwVar}{Variables}
\KwVar{minheap $H$ with entries $\<n, r | len, cpl, n_{prev}, r_{prev}\>$, keys $(len, cpl)$, and compare function $<_{FS}$}

\ForEach{$r_i$ that contains $n_s$}{
	$\la{n_s,r_i} \gets \<n_s,r_i | 0, 0, n_{\varnothing}, r_i\>$\;
	\Enheap $\la{n_s,r_i}$ in $H$ \;
}

\While{$H$ not empty}{
    $\<n_x,r_i | len, cpl, n_w, r_h\> \gets$ \Deheap\;
    \Mark $\la{n_x,r_i}$ as \texttt{final} \;

    \lIf{$n_x$ is $n_t$}{\Break
    }
    \lElse{\ForEach{edge $(n_x, n_y)$} {
    	    \ForEach{road $r_j$ that contains $n_y$} {
	    		\If{$\la{n_y, r_j}$ does not exist} {
	    			$\la{n_y, r_j} \gets \<n_y, r_j | len + L(n_x,n_y), cpl + C(n_y, r_i, r_j), n_x, r_i\>$\;
	    			\Enheap $\la{n_y, r_j}$
	    		}
	    		\ElseIf{$\la{n_y, r_j}$ is not \texttt{final}}{
	    			$ \<n_y, r_j | len', cpl', n_u, r_h\> \gets \la{n_y, r_j}$ \;
	         		\If{ $(len + L(n_x,n_y), cpl + C(n_y, r_i, r_j)) <_{FS} (len', cpl')$ } {
						$\la{n_y, r_j} \gets \<n_y, r_j | len \!+\! L(n_x,n_y), cpl \! + \! C(n_y, r_i, r_j), n_x, r_i\>$\;
						\Update $\la{n_y, r_j}$
		     		}
	    		}
    		}
    	}
    }
}
\Return $(\textit{fsL}, \textit{fsC}) \gets (len,cpl)$ \;
\caption{FastestSimplest}
\label{algo:fs}
\end{algorithm}

%!TEX root = main.tex

\section{Simplest Near-Fastest Route}
\label{sec:near-fastest}

This section studies Problem~\ref{prb:snf}; the solution to
Problem~\ref{prb:sf} is similar and details are omitted. Unlike the case of
finding the simplest fastest or the fastest simplest route, there can exist no
principle of optimality, exactly because the solution to Problem~\ref{prb:snf}
is \emph{not an optimal route} for any definition of optimality. Therefore,
one has to enumerate all routes from source to target, and rely on bounds and
pruning criteria to eliminate sub-routes that cannot be extended to simplest
near-fastest route.

We propose two algorithms, which differ in the way they enumerate paths. The
first, detailed in Section~\ref{sec:SNF-DFS}, is based on depth-first search,
while the second, detailed in Section~\ref{sec:SNF-A}, is inspired by A$^*$
search.

\subsection{DFS-based Traversal}
\label{sec:SNF-DFS}

This section details the SimplestNearFastest-DFS (SNF-DFS) algorithm for
finding the simplest near-fastest route. Its key idea is to enumerate all
routes from source to target by performing a depth-first search, eliminating
in the process routes which are longer than $(1+\epsilon)$ times the fastest
(similar to the algorithm of \cite{BW84} for near-fastest routes), or have
larger complexity than the best found so far.

SNF-DFS requires information about the simplest fastest as well as the fastest
simplest path from any node to the target. To obtain this information, it
invokes two procedures AllFastestSimplest and AllSimplestFastest.

The AllFastestSimplest procedure is a variation of the FastestSimplest
algorithm (Section~\ref{sec:simplest}) that solves the single-source fastest
simplest route problem, i.e., it computes the length and complexity of the
fastest simplest route from a given source to any other node. Only a small
change to the original algorithm is necessary. Recall that when deheaping a
label $\la{n_x, r_i}$, it is marked as final. Observe that when the first
label associated with $n_x$ is deheaped, the algorithm has found the fastest
simplest path from $n_s$ to $n_x$. (This was in fact the termination condition
of Algorithm~\ref{algo:fs}: stop when a label associated with the target is
deheaped.) Therefore, the AllFastestSimplest procedure explicitly marks $n_x$
as visited at its first encounter, and stores the length and complexity of the
current path. The procedure only terminates when the heap empties.

The AllSimplestFastest procedure is derived from the SimplestFastest algorithm
in the same way that AllFastestSimplest is from FastestSimplest, and thus
details are omitted.

Note that there arises a small implementation detail. Recall that the SNF-DFS
algorithm requires the costs all fastest simplest routes \emph{ending at} a
particular node (the target), whereas AllFastestSimplest returns the costs of
all fastest simplest routes \emph{starting from} a particular node. Therefore,
to obtain the appropriate info, SNF-DFS invokes the AllFastestSimplest
procedure using a graph obtained from $G_R$ by inverting the direction of its
edges. The same holds for the invokation of the AllSimplestFastest procedure.

In the following, we assume that the length $\textit{fsL}[\,]$ and complexity
$\textit{fsC}[\,]$ of all fastest simplest routes to the target $n_t$,
and the length $\textit{sfL}[\,]$ and complexity $\textit{sfC}[\,]$ of all
simplest fastest routes to $n_t$, are given.

The SNF-DFS algorithm applies two pruning criteria to avoid examining all
routes from $n_s$ to $n_t$.

\begin{lemma}

Let $\rho$ be a route from $n_s$ to $n_x$. If $L(\rho) + \textit{sfL}[n_x] >
(1+\epsilon)\cdot \textit{sfL}[n_s] $, then any extension of $\rho$ towards
$n_t$ is not a simplest near-fastest route.

\label{lem:prune_length}
\end{lemma}

\alt{}{
\begin{proof}
Any extension of $\rho$ towards $n_t$ must have length at least $L(\rho) +
\textit{sfL}[n_x]$, since $\textit{sfL}[n_x]$ is the shortest length of any
route from $n_x$ to $n_t$. Therefore, the condition of the lemma implies that
no extension of $\rho$ is near-fastest, hence neither simplest near-fastest.
\end{proof}
}

\begin{lemma}

Let $\rho$ be a route from $n_s$ to $n_x$. Further, let $\textit{snfC}^+$ be
an upper bound on the complexity of a simplest near-fastest route from $n_s$
to $n_t$. If $C(\rho) + \textit{fsC}[n_x] > \textit{snfC}^+$, then any
extension of $\rho$ towards $n_t$ is not a simplest near-fastest route.

\label{lem:prune_complexity}
\end{lemma}

\alt{}{
\begin{proof}

Any extension of $\rho$ towards $n_t$ must have complexity at least $C(\rho) +
\textit{sfC}[n_x]$, since $\textit{sfC}[n_x]$ is the lowest complexity of any
route from $n_x$ to $n_t$. Therefore, the condition of the lemma implies that
no extension of $\rho$ has better complexity that an upper bound on the
complexity of the simplest near-fastest route, hence cannot be simplest near-
fastest.
\end{proof}
}

The next lemma computes an upper bound of the complexity of a simplest
near-fastest route.

\begin{lemma}

Let $\rho$ be a route from $n_s$ to $n_x$. If $L(\rho) + \textit{fsL}[n_x]
\leq (1+\epsilon)\cdot \textit{sfL}[n_s] $, then $\textit{snfC}^+ = C(\rho) +
1 + \textit{fsC}[n_x]$ is an upper bound on the complexity of a simplest
near-fastest route.

\label{lem:upper_bound}
\end{lemma}

\alt{}{
\begin{proof}

Consider a simplest extension $\rho'$ of $\rho$ towards $n_t$, i.e., it has
the lowest possible complexity. Observe that its length is $L(\rho') = L(\rho)
+ \textit{fsL}[n_x]$. Hence the condition of the lemma implies that $\rho'$ is
a near-fastest route. So, its complexity is an upper bound on the complexity
of a simplest near-fastest route.

We next show that the complexity of $\rho'$ is at most $C(\rho) + 1 +
\textit{fsC}[n_x]$, which will conclude the proof. Let $n_y$ be the node
following $n_x$ in $\rho'$, and $n_w$ be the node preceding $n_x$ in $\rho$.
Then, the complexity of $\rho'$ is $C(\rho') = C(\rho) + C(n_x, R(n_w, n_x),
R(n_x, n_y)) + \textit{fsC}[n_x]$. Since the turn cost $C(n_x, R(n_w, n_x),
R(n_x, n_y))$ is bounded by 1, any simplest extension of $\rho$ towards $n_t$
has complexity at most $C(\rho) + 1 + \textit{fsC}[n_x]$, which in turn is an
upper bound on the complexity of a simplest near-fastest route.
\end{proof}
}

We are now ready to describe in detail the SNF-DFS algorithm,
whose pseudocode is shown in Algorithm~\ref{algo:snf-dfs}. It
performs a depth-first search on the road network, eliminating routes
according to the two criteria described previously, and computing an upper
bound for the complexity of the fastest near-simplest route.

SNF-DFS uses a stack $S$ to implement depth-first search. At each point in
time, the entries in the stack $S$ form exactly a single route starting from
$n_s$. An entry of $S$ has the form $\< n | len, cpl, n_{prev} \>$, and
corresponds to a route ending at node $n$ with length $len$, complexity $cpl$
and whose second-to-last node is $n_{prev}$.

SNF-DFS marks certain nodes as \texttt{in\_route}, and certain edges as
\texttt{traversed}. Particularly, a node is marked as \texttt{in\_route} if an
entry for this node is currently in the stack. This marking helps avoid cycles
in routes. An edge $(n_a, n_b)$ is marked as \texttt{traversed} if an entry
for $n_a$ is in the stack (not necessarily the last), whereas an entry for
$n_b$ is not in $S$, but was at some previous iteration right above the entry
for $n_a$. This marking helps avoid revisiting routes.

Initially, SNF-DFS invokes the AllSimplestFastest and AllFastestSimplest
procedures (lines 1--2). Then, if the fastest simplest route from $n_s$ to
$n_t$ is near-shortest, i.e., has length less than $(1+\epsilon)\cdot
\textit{sfL}[n_s]$, then it is not only a candidate route but actually the
solution, as there can be no other route with lowest complexity. Hence,
SNF-DFS terminates (lines 3--4).

Otherwise, a candidate route is the fastest simplest route, which is
definitely near-fastest. Therefore, an upper bound on complexity is computed
as $\textit{snfC}^+ = \textit{sfC}[n_s]$ (line 5). The stack is initialized
with an entry for the source node $n_s$ (line 6). Then SNF-DFS proceeds
iteratively until the stack is empty (line 7).

At each iteration the top entry of the stack is examined (but not popped)
(line 8). Let this entry be for node $n_x$ and correspond to a route $\rho$.
If node $n_x$ is not marked as \texttt{in\_route} although it is at the top of
the stack, this means that this is the first time SNF-DFS encounters it (line
9). For this first encounter, the algorithm applies the pruning criterion of
Lemma~\ref{lem:prune_complexity}. If it holds (line 10) then no route that
extends $\rho$ will be examined, and hence the entry is popped from the stack.

Otherwise, if $n_x$ is the target (lines 11--13), $\rho$ constitutes a
candidate solution and its complexity is compared against the best known (line
12). Subsequently, the entry is popped, as there is no need to extend the
current route $\rho$ any farther. If the entry is not popped, node $n_x$ is
marked as \texttt{in\_route}.

If node $n_x$ is not \texttt{in\_route}, SNF-DFS looks for an outgoing edge
$(n_x, n_y)$ such that it is not \texttt{traversed} and $n_y$ is not
\texttt{in\_route} (line 16). If no such edge is found, then all routes, with
no cycles, that extend $\rho$ have been either considered or pruned. Hence the
top entry of the stack is popped (line 18), and $n_x$ is marked as not
\texttt{in\_route} (line 19). Additionally, all outgoind edges of $n_x$ are
marked as not \texttt{traversed} (lines 20--21).

Otherwise, such an outgoing edge $(n_x, n_y)$ is found. Then, the algorithm
checks if the two pruning criteria (Lemmas~\ref{lem:prune_length},
\ref{lem:prune_complexity}) apply (line 22). If either does, then the edge
$(n_x, n_y)$ is marked as \texttt{traversed} (line 27). Otherwise (lines 23--
26), the algorithm checks if Lemma~\ref{lem:upper_bound} applies, and
appropriately updates the complexity bound $\textit{snfC}^+$ if necessary
(line 24). Finally, SNF-DFS creates an entry for node $n_y$ and pushes it in
the stack (line 25), while marking $(n_x, n_y)$ as \texttt{traversed} (line
26).

The actual simplest near-fastest route can be retrieved with standard
backtracking; details are omitted.

\begin{theorem}

The SNF-DFS algorithm correctly finds a simplest near-fastest route from
$n_s$ to $n_t$.

\end{theorem}

\alt{}{
\begin{proof}

We first show that if the pruning criteria were not applied, the algorithm
would enumerate all possible routes from $n_s$ to $n_t$. This is true, because
SNF-DFS would perform a depth-first traversal constructing each time an
acyclic route consisting of possibly all edges until $n_t$ is reached (line
11). The marking on edges guarantees that when the algorithm backtracks
(performs a pop operation), a different route is followed. Eventually, when
the stack empties the algorithm would have constructed all routes from $n_s$
to $n_t$.

We finally argue that all pruned routes cannot be sub-routes of a simplest
near-fastest route. This holds because pruning is performed based on
Lemmas~\ref{lem:prune_length} and \ref{lem:prune_complexity}, and the bound of
Lemma~\ref{lem:upper_bound}.
\end{proof}
}

\stitle{Analysis.}
The complexities of AllSimplestFastest and AllFastestSimplest are the same as
those of SimplestFastest and FastestSimplest, respectively, namely $O(\delta^2
|E| + \delta |V|\log|V|)$ amortized time and $O(\delta |V|)$ space.

Let $L(\rho^{SF})$ denote the length of the fastest route, and $\Delta d$ the
smallest distance of any edge. At any time the stack of SimplestNearFastest
corresponds to a sub-route of some near-fastest route. The number of edges in
a near-fastest route can be at most $(1+\epsilon)L(\rho^{SF})/\Delta d$ (but
not more than $|E|$).  Therefore, the space complexity of the road network
traversal is $O((1+\epsilon)L(\rho^{SF})/\Delta d) = O(|E|)$, since at each
time a single route is maintained in the stack.

In the worst case, the traversal may examine all possible routes from $n_s$ to
$n_t$ having $(1+\epsilon)L(\rho^{SF})/\Delta d$ edges. The number of such
routes is $k = \binom{|E|}{(1+\epsilon)L(\rho^{SF})/\Delta d}$; in practice
this is a much smaller number. The number of push or pop operations is in the
worst case equal to the total length of all possible near-fastest routes from
source to target. Since there can be $k$ such routes, the time complexity is
$O(k(1+\epsilon)L(\rho^{SF})/\Delta d)$.

Overall, the time complexity of SNF-DFS is $O(\delta^2 |E| + \delta |V|\log|V|
+ k(1+\epsilon)L(\rho^{SF})/\Delta d)$ amortized, while its space complexity
is $O(\delta |V| + (1+\epsilon)L(\rho^{SF})/\Delta d)$.

\begin{algorithm}[t]
\scriptsize
\SetInd{0.3em}{0.8em}
\SetCommentSty{textsf}
\SetArgSty{textrm}
\DontPrintSemicolon
\SetKw{Continue}{continue}
\SetKw{Break}{break}
\SetKw{Find}{find}
\SetKw{Push}{push}
\SetKw{Pop}{pop}
\SetKw{ReadTop}{top}
\SetKw{MyAnd}{and}
\SetKw{Mark}{mark}
\KwIn{road network $G_R$; mapping $C$; source $n_s$; target $n_t$; value $\epsilon$}
\KwOut{length \textit{snfL} and complexity \textit{snfC} of simplest near-fastest route from $n_s$ to $n_t$}
\SetKwInput{KwVar}{Variables}
\KwVar{stack $S$ with entries $\< n | len, cpl, n_{prev} \>$}

$(\textit{sfL}[\,],\textit{sfC}[\,]) \gets \mathrm{AllSimplestFastest}(G_R,C,n_t)$ \;
$(\textit{fsL}[\,],\textit{fsC}[\,]) \gets \mathrm{AllFastestSimplest}(G_R,C,n_t)$ \;

\If{ $\textit{fsL}[n_s] \leq (1+\epsilon)\cdot \textit{sfL}[n_s]$} {
     \Return $(\textit{snfL}, \textit{snfC}) \gets (\textit{fsL}[n_s], \textit{fsC}[n_s])$\;
} 

$(\textit{snfL}^+, \textit{snfC}^+) \gets (\textit{sfL}[n_s], \textit{sfC}[n_s])$ \;

\Push $(n_s | 0, 0, n_{\varnothing})$ \;

\While{$S$ not empty}{
     $\< n_x | len, cpl, n_w \> \gets $ \ReadTop \;
     
     \If{$n_x$ not \texttt{in\_route} }
     {
          \lIf{ $cpl + \textit{fsC}[n_x] > \textit{snfC}^+$ } {\Pop
          }
          \ElseIf{$n_x$ is $n_t$}{
               \lIf{$cpl < \textit{snfC}^+$} {
          	    $(\textit{snfL}^+, \textit{snfC}^+) \gets (len, cpl)$
               }
               \Pop
          }
          \lElse{
               \Mark $n_x$ as \texttt{in\_route}
          }
     }

     \Else{
	     \Find an outgoing edge $(n_x, n_y)$ that is not \texttt{traversed} and $n_y$ is not \texttt{in\_route} \;
          \If {no such edge is found} {
               \Pop\;
               \Mark $n_x$ as not \texttt{in\_route}\;
               \ForEach {outgoing edge $(n_x, n_y)$} {
                    \Mark $(n_x,n_y)$ as not \texttt{traversed}\;
               }        
          }
     	\ElseIf{$len+ L(n_x, n_y) + \textit{sfL}[n_y] \leq (1\!+\!\epsilon)\!\cdot\! \textit{sfL}[n_s]$ \MyAnd 
     		$cpl + C(e_{wx},e_{xy},n_x) + \textit{fsC}[n_y] < \textit{snfC}^+ $ } {
               
               \If{$len+ L(n_x, n_y) + \textit{fsL}[n_y] \leq (1\!+\!\epsilon)\!\cdot\! \textit{sfL}[n_s]$ } {
                    $(\textit{snfL}^+, \textit{snfC}^+) \gets (len+ L(n_x, n_y) + \textit{fsL}[n_y], cpl + C(e_{wx},e_{xy},n_x) + 1 + \textit{fsC}[n_y])$\;
               }
     		\Push $\<n_y | len+ L(n_x, n_y), cpl + C(e_{wx},e_{xy},n_x), n_x\>$\;
               \Mark $(n_x,n_y)$ as \texttt{traversed}\;
     	}
          \lElse {
               \Mark $(n_x,n_y)$ as \texttt{traversed}\;
          }    
     }
}
\Return $(\textit{snfL}, \textit{snfC}) \gets (\textit{snfL}^+, \textit{snfC}^+)$\;
\caption{SimplestNearFastest-DFS}
\label{algo:snf-dfs}
\end{algorithm}

\stitle{Discussion.}
The running time of SNF-DFS depends on large part on the two procedures
AllSimplestFastest and AllFastestSimplest. In the following, we discuss a
variant of the algorithm that does not invoke these procedures. The key idea
is to relax the requirement for explicit calculation of the length and
complexity of all simplest fastest and fastest simplest routes, and instead
require a method for calculating their lower and upper bounds. Such a method
can be straightforwardly adapted from landmark-based techniques, e.g.,
\cite{GH05}. Note that in the extreme case, no bounds are necessary.
\eat{
Therefore, in the following, we assume that a lower $\textit{fsL}^-[\,]$ and
an upper $\textit{fsL}^+[\,]$ bound on the length, and a lower
$\textit{fsC}^-[\,]$ and an upper $\textit{fsC}^+[\,]$ bound on the complexity
of all fastest simplest routes  to the target $n_t$ are known, or can be
computed in constant time. Similarly, there exist bounds,
$\textit{sfL}^-[\,]$, $\textit{sfL}^+[\,]$, $\textit{sfC}^-[\,]$,
$\textit{sfC}^+[\,]$ for all simplest fastest routes.
}
Clearly, the pruning criteria of Lemmas~\ref{lem:prune_length} and
\ref{lem:prune_complexity} can be straightforwardly adapted to use bounds
instead; note that their pruning power is reduced. Similarly,
Lemma~\ref{lem:upper_bound} can also be adapted, which results however in a
less tight upper bound. Details are omitted.

\subsection{A$^*$-based Traversal}
\label{sec:SNF-A}

This section describes the SimplestNearFastest-A$^*$ (SNF-A$^*$) algorithm for
finding a simplest near-fastest route, which is inspired by A$^*$ search. The
key idea is to use bounds on the complexity in order to guide the search
towards the simplest among the near-fastest routes.

Similar to the dfs-like algorithm, SNF-A$^*$ applies
Lemmas~\ref{lem:prune_length}, \ref{lem:prune_complexity} to prune unpromising
routes, and Lemma~\ref{lem:upper_bound} to compute an upper bound on the
complexity of a simplest near-fastest route. On the other hand, contrary to
the dfs-like algorithm, SNF-A$^*$ terminates when it enounters the target node
for the first time, because it can guarantee that all unexamined routes have
more complexity.

The SNF-A$^*$ algorithm uses a heap to guide the search, containing node
labels. An important difference with respect to the methods of
Section~\ref{sec:simplest}, is that to guarantee correctness, there may be
multiple labels per node, each corresponding to different routes from the
source to that node. The reason is that there is no principle of optimality
for near-fastest routes. Still, labels belonging to certain routes can be
eliminated, as the following lemma suggests.

\begin{lemma}

Let $\rho$, $\rho'$ be two routes from $n_s$ to $n_x$. If $L(\rho') > L(\rho)$
and $ C(\rho') > C(\rho) + 1$, then $\rho'$ cannot be a sub-route of a
simplest near-fastest route from $n_s$ to any $n_t$.

\label{lem:prune_dominance}
\end{lemma}

\alt{}{
\begin{proof}

Let $n_w$ (resp. $n_{w'}$) be the second-to-last node of route $\rho$ (resp.
$\rho'$). We prove by contradiction. Assume that $\rho'$ is a sub-route of a
simplest near-fastest route $\rho^{FS'}$. Let $\rho_x$ be the sub-route of
$\rho^{FS'}$ starting from node $n_x$ and ending at $n_t$, an let $n_y$ be its
second node, after $n_x$. Then, $L(\rho^{FS'}) = L(\rho') + L(\rho_x)$, and
$C(\rho^{FS'}) = C(\rho') + C(n_x, R(n_{w'}, n_x), R(n_x, n_y)) + C(\rho_x)$.
Since in the best case, a turn cost can be zero, we
have that $C(\rho^{FS'}) \geq C(\rho') + C(\rho_x)$.

Now consider route $\rho^{FS} = \rho \rho_x$, where $L(\rho^{FS}) = L(\rho) +
L(\rho_x)$, and $C(\rho^{FS}) = C(\rho) + C(n_x, R(n_{w}, n_x), R(n_x, n_y)) +
C(\rho_x)$. Since in the worst case, a turn cost can be one, we have that
$C(\rho^{FS}) \leq C(\rho) + 1 + C(\rho_x)$. From the conditions of the lemma,
we derive that $L(\rho^{FS}) < L(\rho^{FS'})$ and $C(\rho^{FS}) <
C(\rho^{FS'})$. This implies that $\rho^{FS}$ is near-fastest, as it has
length less than a near-fastest route. Moreover, it has less complexity than
$\rho^{FS'}$, which is a contradiction as $\rho^{FS'}$ is simplest
near-fastest.
\end{proof}
}

The set of labels for a node $n_x$ is denoted by $\Lambda(n_x)$. Let $\lae$
(resp. $\lae'$) be the label corresponding to a route $\rho$ (resp. $\rho'$)
ending at node $n_x$. If the conditions of Lemma~\ref{lem:prune_dominance}
hold for $\rho$ and $\rho'$, we write $\lae \prec \lae'$.  Clearly, there is
no need to keep a label $\lae' \in \Lambda(n_x)$ if there is another label
$\lae \in \Lambda(n_x)$ such that $\lae \prec \lae'$.

An important difference to the label-setting method for Problem~\ref{prb:fs}
is that a heap entry (label) $\< n | len, cpl, n_{prev} \>$ in SNF-A$^*$ is
sorted according to the  FS-shorter total order (see
Section~\ref{sec:simplest}) on pair $(len+ \textit{fsL}[n],
cpl+\textit{fsC}[n])$, as it would in A$^*$ search.

The pseudocode of SNF-A$^*$ is shown in Algorithm~\ref{algo:snf-a}. Initially,
it invokes the AllSimplestFastest and AllFastestSimplest procedures to obtain
arrays $\textit{fsL}[\,]$, $\textit{fsC}[\,]$, $\textit{sfL}[\,]$, and
$\textit{sfC}[\,]$ (lines 1--2). Subsequently, if the fastest simplest route
from $n_s$ to $n_t$ is near-shortest, it is the solution, and hence, SNF-DFS
terminates (lines 3--4). Otherwise, a candidate route is the fastest simplest
route, which is definitely near-fastest. Therefore, an upper bound on
complexity is computed as $\textit{snfC}^+ = \textit{sfC}[n_s]$ (line 5).

\begin{algorithm}[t]
\scriptsize
\SetInd{0.3em}{0.8em}
\SetCommentSty{textsf}
\SetArgSty{textrm}
\DontPrintSemicolon
\SetKw{Break}{break}
\SetKw{Continue}{continue}
\SetKw{Enheap}{enheap}
\SetKw{Deheap}{deheap}
\SetKw{Retrieve}{retrieve}
\SetKw{Mark}{mark}
\SetKw{MyAnd}{and}
\SetKw{Remove}{remove}
\KwIn{road network $G_R$; mapping $C$; source $n_s$; target $n_t$; value $\epsilon$}
\KwOut{length \textit{snfL} and complexity \textit{snfC} of simplest near-fastest route from $n_s$ to $n_t$}
\SetKwInput{KwVar}{Variables}
\KwVar{minheap $H$ with entries $\<n | len, cpl, n_{prev}\>$, keys $(len\!+\!fsL[n], cpl\!+\!fsC[n])$, compare function $<_{FS}$}

$(\textit{sfL}[\,],\textit{sfC}[\,]) \gets \mathrm{AllSimplestFastest}(G_R,C,n_t)$ \;
$(\textit{fsL}[\,],\textit{fsC}[\,]) \gets \mathrm{AllFastestSimplest}(G_R,C,n_t)$ \;

\If{ $\textit{fsL}[n_s] \leq (1+\epsilon)\cdot \textit{sfL}[n_s]$} {
     \Return $(\textit{snfL}, \textit{snfC}) \gets (\textit{fsL}[n_s], \textit{fsC}[n_s])$\;
} 

$(\textit{snfL}^+, \textit{snfC}^+) \gets (\textit{sfL}[n_s], \textit{sfC}[n_s])$ \;

\Enheap $\<n_s| 0, 0, n_{\varnothing}\>$ in $H$ \;

\While{$H$ not empty}{
     $\<n_x | len, cpl, n_w\> \gets$ \Deheap\;
     \If{$n_x$ is $n_t$}{
          $(\textit{snfL}^+, \textit{snfC}^+) \gets (len, cpl)$ \;
          \Break\;
     }
     \lElse{\ForEach{edge $(n_x, n_y)$}{
               $\lae \gets \<n_y | len+ L(n_x, n_y), cpl + C(e_{wx},e_{xy},n_x), n_x\>$ \;
               $pruned \gets \texttt{false}$\;
               \ForEach{entry $\lae' \in \Lambda(n_y)$} {
                    \lIf{ $\lae \prec \lae'$ }{\Remove $\lae'$}
                    \lElseIf{$\lae' \prec \lae$}{$pruned \gets \texttt{true}$}
               }
               \If{not $pruned$ \MyAnd $len+ L(n_x, n_y) + \textit{sfL}[n_y] \leq (1\!+\!\epsilon)\!\cdot\! \textit{sfL}[n_s]$ \MyAnd 
               $cpl + C(e_{wx},e_{xy},n_x) + \textit{fsC}[n_y] < \textit{snfC}^+ $ }
               {              
                    \If{$len+ L(n_x, n_y) + \textit{fsL}[n_y] \leq (1\!+\!\epsilon)\!\cdot\! \textit{sfL}[n_s]$ }
                    {
                         $(\textit{snfL}^+, \textit{snfC}^+) \gets (len+ L(n_x, n_y) + \textit{fsL}[n_y], cpl + C(e_{wx},e_{xy},n_x) + 1 + \textit{fsC}[n_y])$
                    }
                    \Enheap $\lae$
               }
          }
     }          
}
\Return $(\textit{snfL}, \textit{snfC}) \gets (\textit{snfL}^+, \textit{snfC}^+)$ \;
\caption{SimplestNearFastest-A$^*$}
\label{algo:snf-a}
\end{algorithm}

The heap is initialized with an entry for the source node $n_s$ (line 6). Then
SNF-A$^*$ proceeds iteratively until the heap is empty (line 7). Let $\<n_x |
len, cpl, n_w\>$ be the deheaped label at some iteration (line 8). If $n_x$ is
the target, the algorithm terminates (lines 9--11). The reason is that because
of the order in the heap, all remaining labels correspond to routes, which
when extended via the simplest route to the target, have larger complexity.
Hence, Lemma~\ref{lem:prune_complexity} applies to them.

If $n_x$ is not the target, each outgoing edge $(n_x, n_y)$ is examined (line
12), and a label $\lae$ for the route to $n_y$ is created (line 13).
Subsequently, each other label $\lae'$ regarding node $n_y$ is considered
(lines 15--17). In particular, the algorithm applies
Lemma~\ref{lem:prune_dominance} for the routes of labels $\lae$ and $\lae'$, removing labels
if necessary.

If the route for label $\lae$ survives, then the pruning criteria of
Lemmas~\ref{lem:prune_length} and \ref{lem:prune_complexity} are applied (line
18). If the label still survives, then Lemma~\ref{lem:upper_bound} is applied
to compute an upper bound on the complexity of a solution (lines 19--20). Finally, the
surviving label $\lae$ is enheaped (line 21).

As before, the actual simplest near-fastest route can be retrieved with
standard backtracking; details are omitted.

\begin{theorem}

The SNF-A$^*$ algorithm correctly finds a simplest near-fastest route from
$n_s$ to $n_t$.

\end{theorem}

\alt{}{
\begin{proof}

We first show that if the pruning criteria and the termination condition were
not applied, the algorithm would enumerate all possible routes from $n_s$.
This is true because when a label is deheaped for node $n_x$, a route is
identified, which is subsequently extended by considering all the neighbors of
$n_x$. Note that multiple labels for node $n_x$ might be deheaped,
corresponding to different routes, possibly with cycles. The fact that the
heap entries are sorted by the FS-shorter order, i.e., primarily by complexity
and secondarily by length, and the fact that a route with cycles is always not
FS-shorter than its acyclic counterpart, ensures that the algorithm does not
fall into an endless loop traversing a cycle, and will eventually examine all
routes.

Next, we show that all pruned labels correspond to routes that cannot be 
sub-routes of a simplest near-fastest route. This is true, because pruning is
performed based on Lemmas~\ref{lem:prune_length}, \ref{lem:prune_complexity}
and \ref{lem:prune_dominance} and the bound of Lemma~\ref{lem:upper_bound}.

Finally, we show that when SNF-A$^*$ terminates (line 11), a simplest 
near-fastest route is identified. Let $\rho$ denote the route that corresponds to
the deheaped label $\lae$ for the target $n_t$. Observe that $\rho$ is 
near-fastest, because otherwise its label $\lae$ would not be enheaped at line 21
(pruned by Lemmma~\ref{lem:prune_length}). We finally argue that $\rho$ has
the lowest complexity among all near-fastest routes. This holds due to the 
FS-shorter order of the heap. All other routes to $n_t$ have complexity not less
than $\rho$'s.
\end{proof}
}

\stitle{Analysis.}
AllSimplestFastest and AllFastestSimplest require $O(\delta^2 |E| + \delta
|V|\log|V|)$ amortized time and $O(\delta |V|)$ space.
In the worst case, the algorithm may examine all $k =
\binom{|E|}{(1+\epsilon)L(\rho^{SF})/\Delta d}$ possible routes from $n_s$ to
$n_t$ having at most $(1+\epsilon)L(\rho^{SF})/\Delta d$ edges. Each node of
the road network may be assigned up to $k$ labels, one per possible route. The
number of enheap and deheap operations equals the number of labels $k|V|$.
Furthermore, the number of update operations is equal to $k^2$ per edge, for a
total of $k^2|E|$. Assuming a Fibonacci heap, the time complexity of the
traversal is $O(k^2 |E| + k |V|\log|V|)$ amortized. Moreover, since the heap
may contain an entry for each label, the space complexity is $O(k |V|)$.
Overall, the time complexity of SNF-A$^*$ is $O(\delta^2 |E| + \delta
|V|\log|V| + k^2 |E| + k |V|\log|V|)$ amortized, while its space complexity is
$O(\delta |V| + k |V|)$.

\stitle{Discussion.} Similarly to the case of SNF-DFS, the invocation of the AllSimplestFastest and AllFastestSimplest procedures is not necessary for SNF-A$^*$.

%!TEX root = main.tex

\section{Experimental Evaluation}
\label{sec:exps}

This section, presents an experiment evaluation of our methodology for Problems~\ref{prb:fs}--\ref{prb:snf}. Section~\ref{sec:setup} details the setup of our analysis. Section~\ref{sec:comparison} qualitatively compares the proposed methods, and Section~\ref{sec:scalability} studies the scalability.

\begin{figure}[b]
\vspace{-10pt}
\centering
\includegraphics[width=5cm]{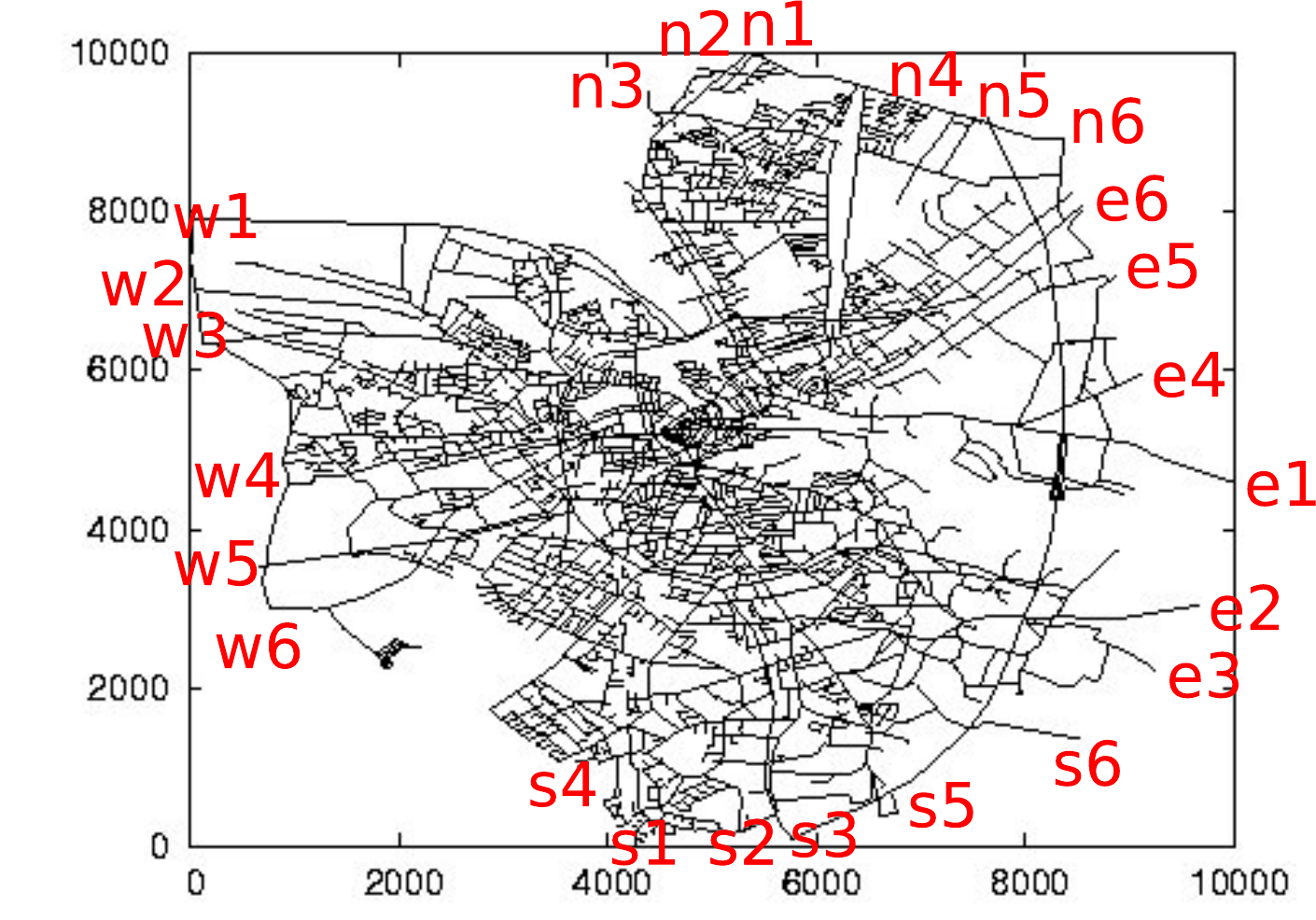}
\vspace{-5pt}
\caption{The OLB road network and its 24 entrances/exits.}
\label{fig:olb_map}
\end{figure}

\begin{figure}
\centering
\begin{tabular}{cc}
\includegraphics[width=0.25\columnwidth]{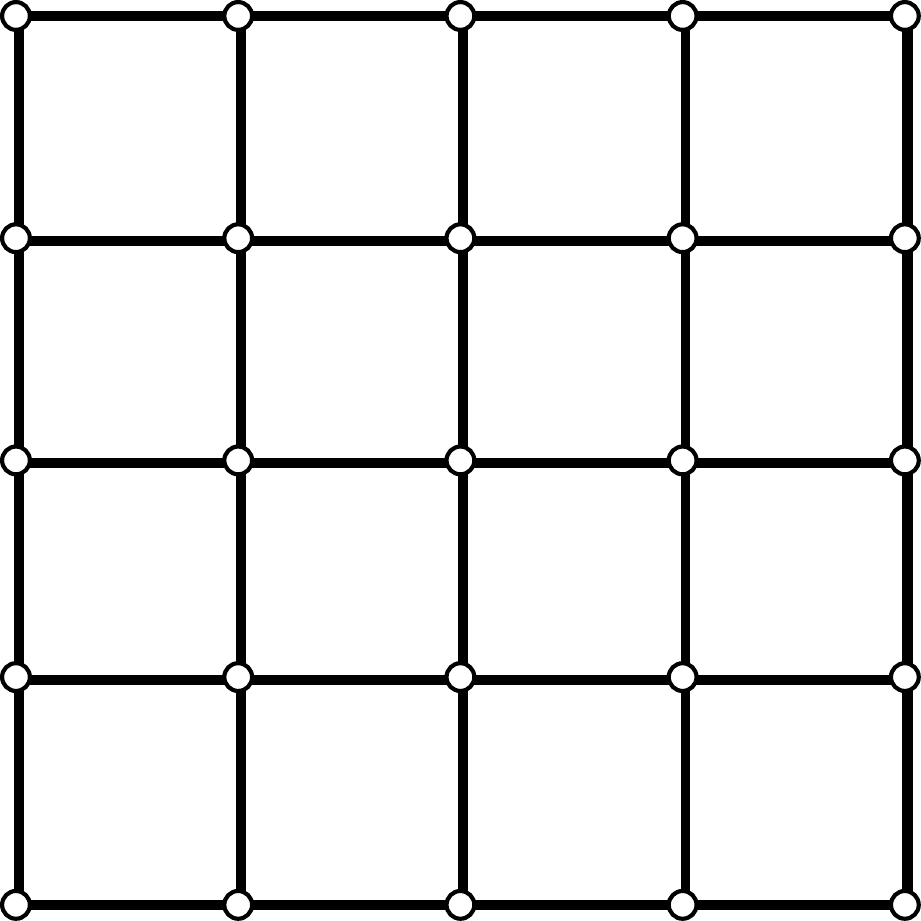} \hspace{10pt}
&\hspace{10pt}\includegraphics[width=0.25\columnwidth]{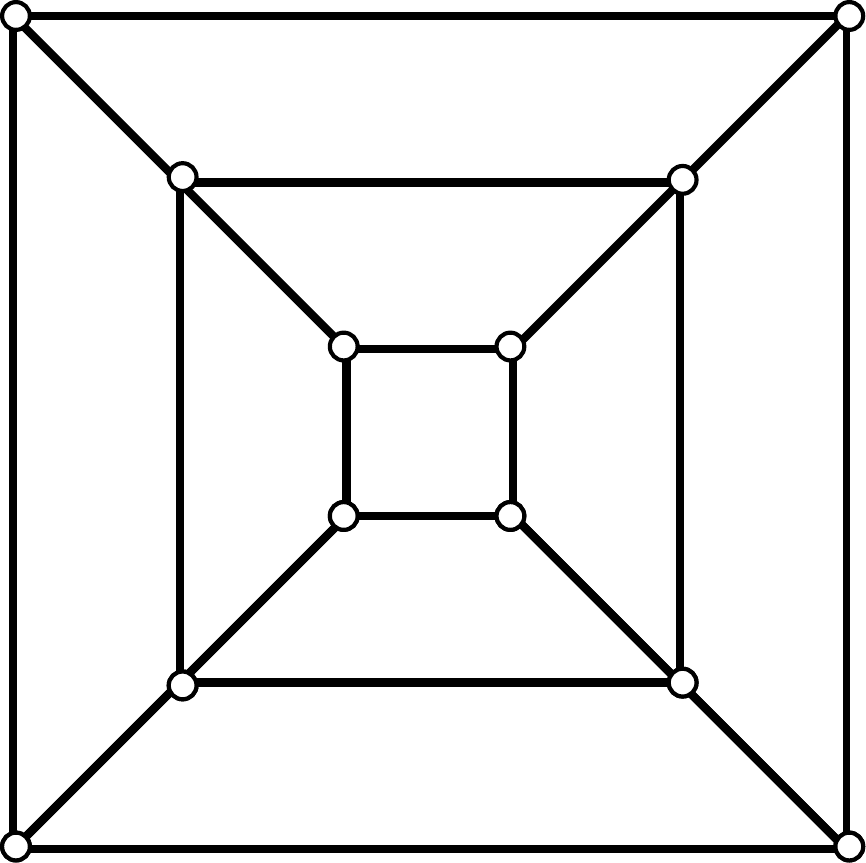}\\
(a) &(b)\\
\end{tabular}
\vspace{-10pt}
\caption{Examples of backbone networks: (a) grid-based with 10 roads, (b) ring-based with 16 roads.}
\vspace{-10pt}
\label{fig:backbone}
\end{figure}

\subsection{Setup}
\label{sec:setup}
Our experimental analysis involves both real and synthetic road networks. We use the real road networks of the following cities taken from OpenStreetMap\eat{\footnote{\url{http://www.openstreetmap.org/}}}: Oldeburg (OLB), Berlin (BER), Vienna (VIE) and Athens (ATH), containing $1,672$ roads and $2,383$ intersections, $15,246$ roads and $25,321$ intersections, $20,224$ roads and $27,563$ intersections, and $76,896$ roads and  $108,156$ intersections, respectively. The weighted average degree of an intersection in these road networks is $2.09$, $2.15$, $2.17$ and $2.19$, respectively.

To study the scalability of our methodology we also generated synthetic road networks by populating the OLB road network. The idea is the following. In an attempt to capture the structure of a real network, a synthetic road network is defined as a set of neighborhoods connected to each other through a backbone road network. OLB is used to capture the internal road network of a neighborhood. Figure~\ref{fig:olb_map} pictures the 24 intersections used to enter/exit the internal road network of a neighborhood from/to the backbone.

Finally, to construct a backbone network we consider two different topologies. The grid-based topology of degree $\tau$ is constructed by $2\tau$ roads, $\tau^2$ intersections, and defines $(\tau-1)^2$ neighborhoods. On the other hand, a ring-based topology of degree $\tau$ is constructed by $4(\tau +1)$ roads, $4\tau$ intersections, and defines $4(\tau-1) + 1$ neighborhoods. 
Figure~\ref{fig:backbone}(a) and (b) show an example of a grid-based and a ring-based backbone road network of degrees 5, and 3, respectively. The grid-based backbone consists of 10 roads connected through 25 intersections and defines 16 neighborhoods, while the ring-based backbone consists of 16 roads connected through 12 intersections and defines 9 neighborhoods. 
%In our experiments we vary the number of neighborhoods included in a synthetic road network inside $\{1,4,9,16\}$ when a grid-based backbone is used, and inside $\{1,5,9,13\}$ in case of a ring-based backbone network.

To assess the performance of the routing methods, we measure their average response time and the average number of routes examined over $1,000$ queries. Finally, in case of the simplest near-fastest and the fastest near-simplest route problems, we test the methods varying $\epsilon$ inside $\{0.01,0.05,$ $0.1,0.2,0.3\}$.

\subsection{Comparison of Routing Methods}
\label{sec:comparison}
\begin{table*}[ht]
\scriptsize
\centering
\caption{Real road networks: performance analysis for solving Problems~\ref{prb:fs} and \ref{prb:sf}.}
\vspace{2pt}
\label{tab:comparison_sf-fs}
\begin{tabular}{ccccccc}
\toprule
 &\multicolumn{2}{c}{BSL} &\multicolumn{2}{c}{FS} &\multicolumn{2}{c}{SF}\\
\textbf{road} &\textbf{Response} &\textbf{Routes} &\textbf{Response} &\textbf{Routes} &\textbf{Response} &\textbf{Routes}\\
\textbf{network} &\textbf{time (sec)} &\textbf{examined} &\textbf{time (sec)} &\textbf{examined} &\textbf{time (sec)} &\textbf{examined}\\
\midrule
OLB &$68.7$ &$121,236,000$ &$0.003$ &$2286.82$ &$0.003$ &$2418.35$\\
BER &$-$ &$-$ &$0.055$ &$27226.3$ &$0.040$ &$27611.7$\\
VIE &$-$ &$-$ &$0.057$ &$29301.8$ &$0.042$ &$29250.8$\\
ATH &$-$ &$-$ &$0.346$ &$117,973$ &$0.207$ &$120,329$\\
\bottomrule
\end{tabular}
\end{table*}
The first set of experiments involves the OLB, BER, VIE, and ATH real road networks with the purpose of identifying the best method for each of the problems at hand. 

Table~\ref{tab:comparison_sf-fs} demonstrates the results for the fastest simplest and the simplest fastest route problems. We first observe that FS outperforms BSL by several orders of magnitude. In fact, we managed to execute BSL only on the smallest road network (OLB) due to its extremely high response time. This is expected as BSL needs to enumerate an enormous number of routes to identify the final answer. On the other hand, we observe that FS, SF identify the corresponding routes in less than half a second for all real networks.

\begin{figure*}[!ht]
\begin{center}
\begin{tabular}{cccc}
\hspace{-0.5cm}
\includegraphics[width=0.245\linewidth]{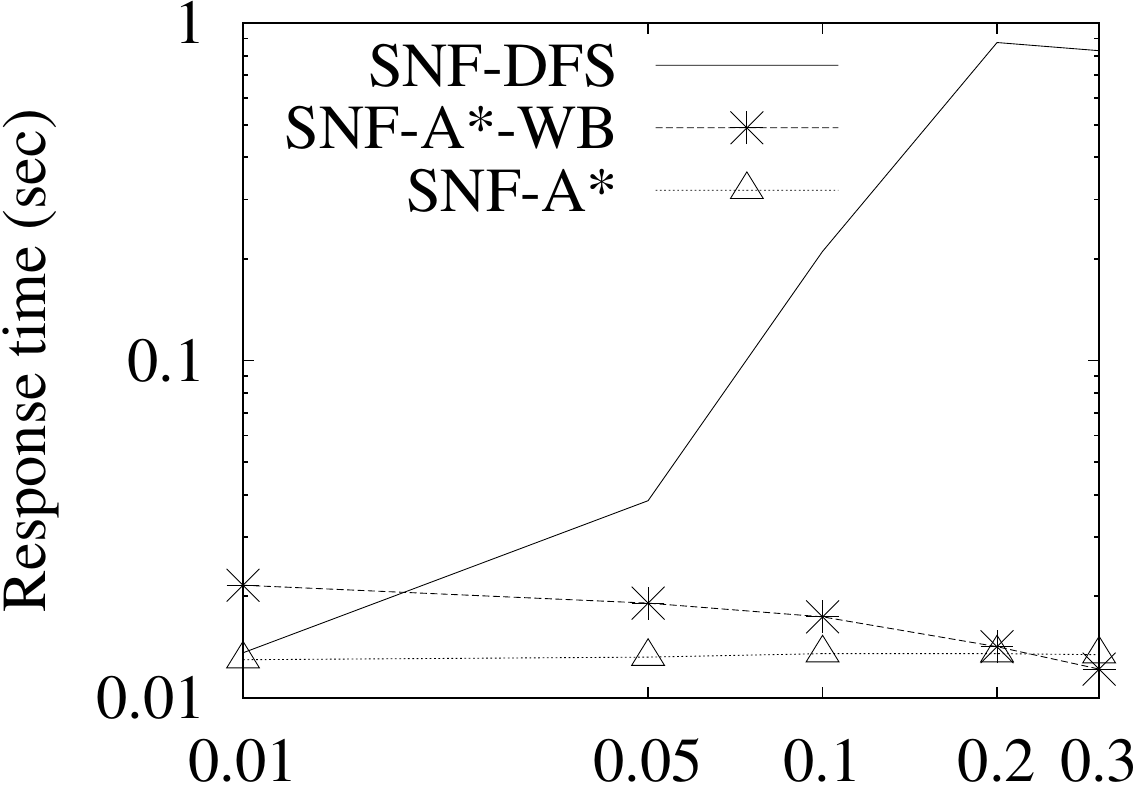}
&\hspace{-0.2cm}\includegraphics[width=0.245\linewidth]{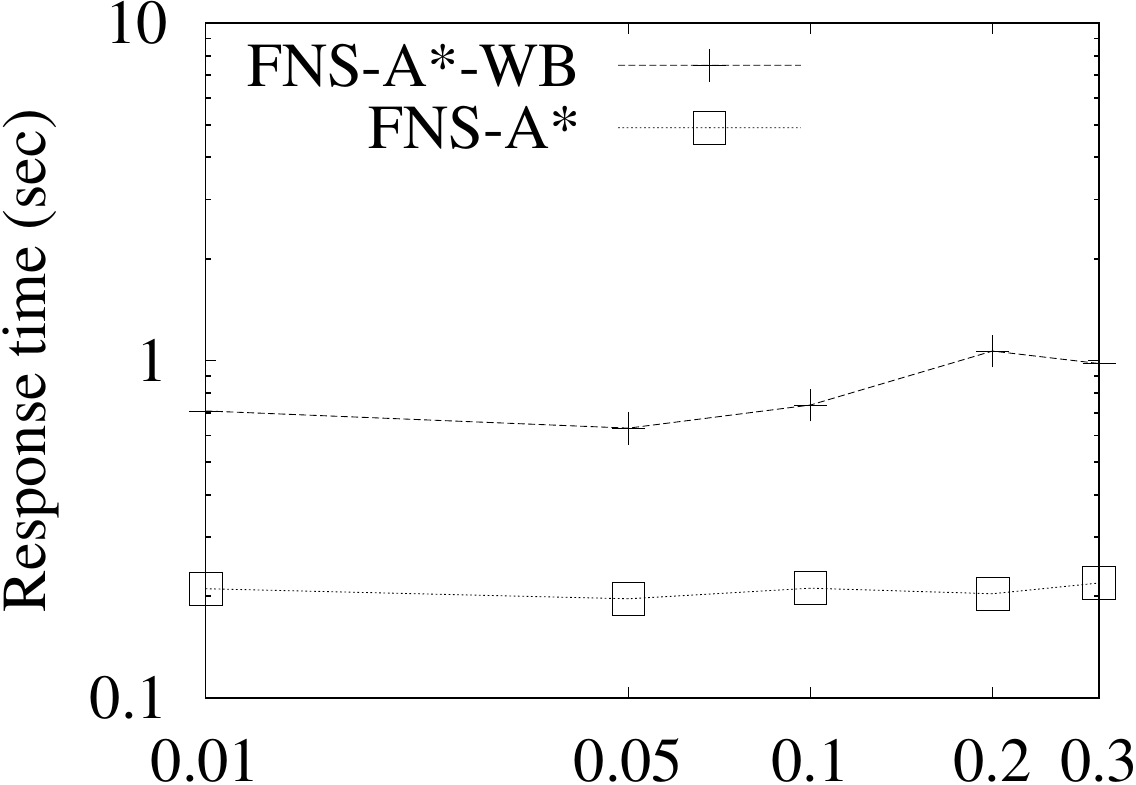}
&\hspace{-0.2cm}\includegraphics[width=0.245\linewidth]{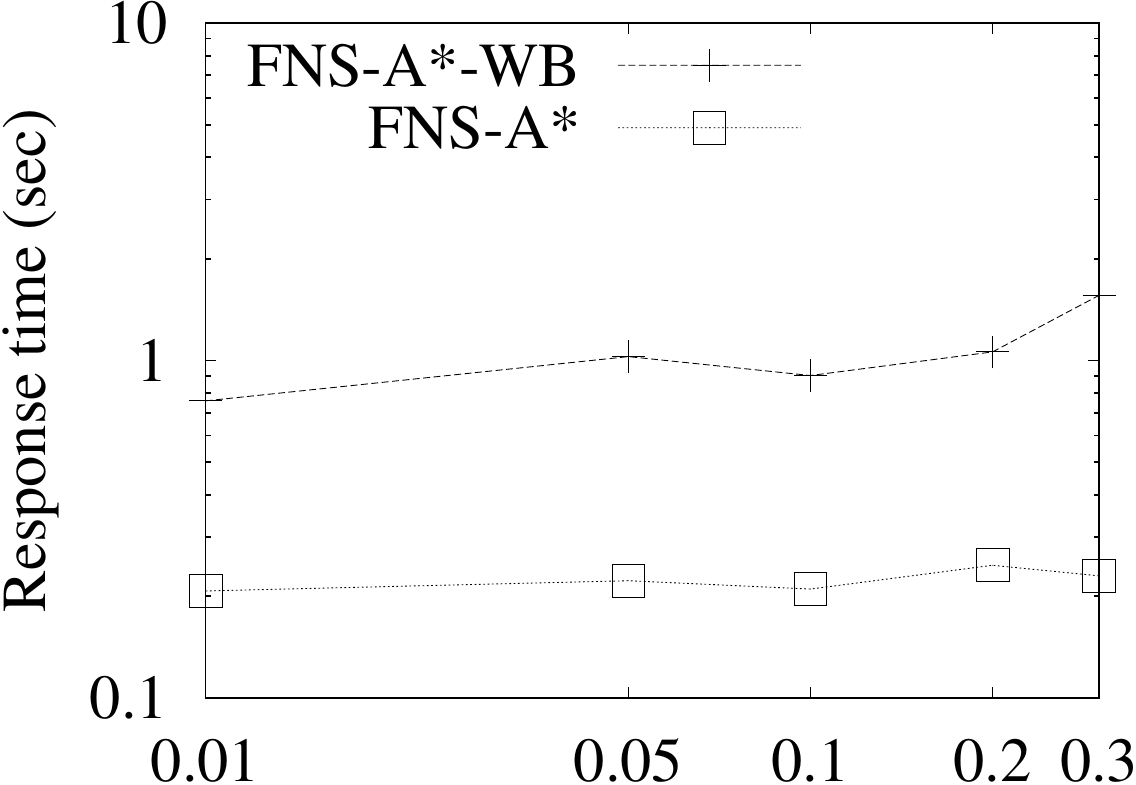}
&\hspace{-0.2cm}\includegraphics[width=0.245\linewidth]{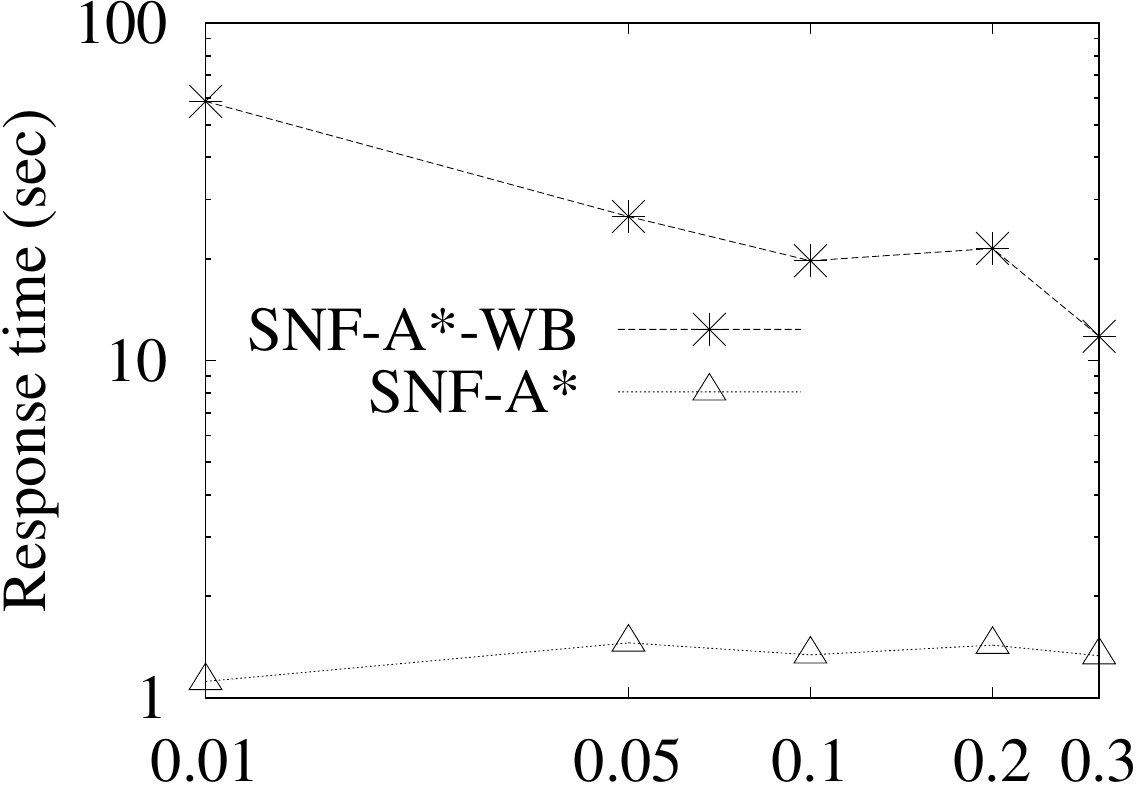}\\
$\epsilon$ &$\epsilon$ &$\epsilon$ &$\epsilon$\\
(a) OLB
&(b) BER
&(c) VIE
&(d) ATH\\
\end{tabular}
\vspace{-10pt}
\caption{Real road networks: performance analysis for solving Problems~\ref{prb:fns} and \ref{prb:snf}.}
\vspace{-10pt}
\label{fig:comparison_snf-fns}
\end{center}
\end{figure*}

Finally, we investigate which is the best method for the fastest near-simplest and the simplest near-fastest route problems. Note that for the purpose of this experiment we include two additional methods termed FNS-A$^*$-WB and SNF-A$^*$-WB. These algorithms follow the same principle as FNS-A$^*$ and SNF-A$^*$ respectively, without however invoking the AllFastestSimplest and AllSimplestFastest procedures (equivalently they assume $sfL[n] = sfC[n] = fsL[n] = fsC[n] = 0$ for any node $n$). In addition, note that because of their high response time, we were able to execute FNS-DFS and SNF-DFS only on the smallest road network, OLB. Figure~\ref{fig:comparison_snf-fns} clearly shows that FNS-A$^*$ and SNF-A$^*$ are the dominant methods for the problems at hand. In fact with the exception of the smallest road network, OLB, they outperform their competitors by at least one order of magnitude. The superiority of FNS-A$^*$ (SNF-A$^*$) over FNS-A$^*$-WB (SNF-A$^*$-WB) supports our decision to invoke the AllFastestSimplest and AllSimplestFastest procedures before the actual search takes place. 

We also observe that as $\epsilon$ increases, the response time of the methods that solve simplest near-fastest route problem decreases. Specifically, the response time of SNF-A$^*$-WB always decreases while the time of SNF-A$^*$ first increases and after $\epsilon = 0.1$ or $\epsilon = 0.2$ it drops. Note that this trend is also followed by the average number of routes examined by the methods. The reason for is that the larger $\epsilon$ is, the more routes have acceptable length and thus need to be examined. At the same time, however, it is more likely to early identify a candidate answer, which can enhance the pruning mechanism and thus accelerate the query evaluation.

\subsection{Scalability Tests}
\label{sec:scalability}
\begin{figure*}[!ht]
\begin{center}
\begin{tabular}{cccc}
\hspace{-0.5cm}
\includegraphics[width=0.245\linewidth]{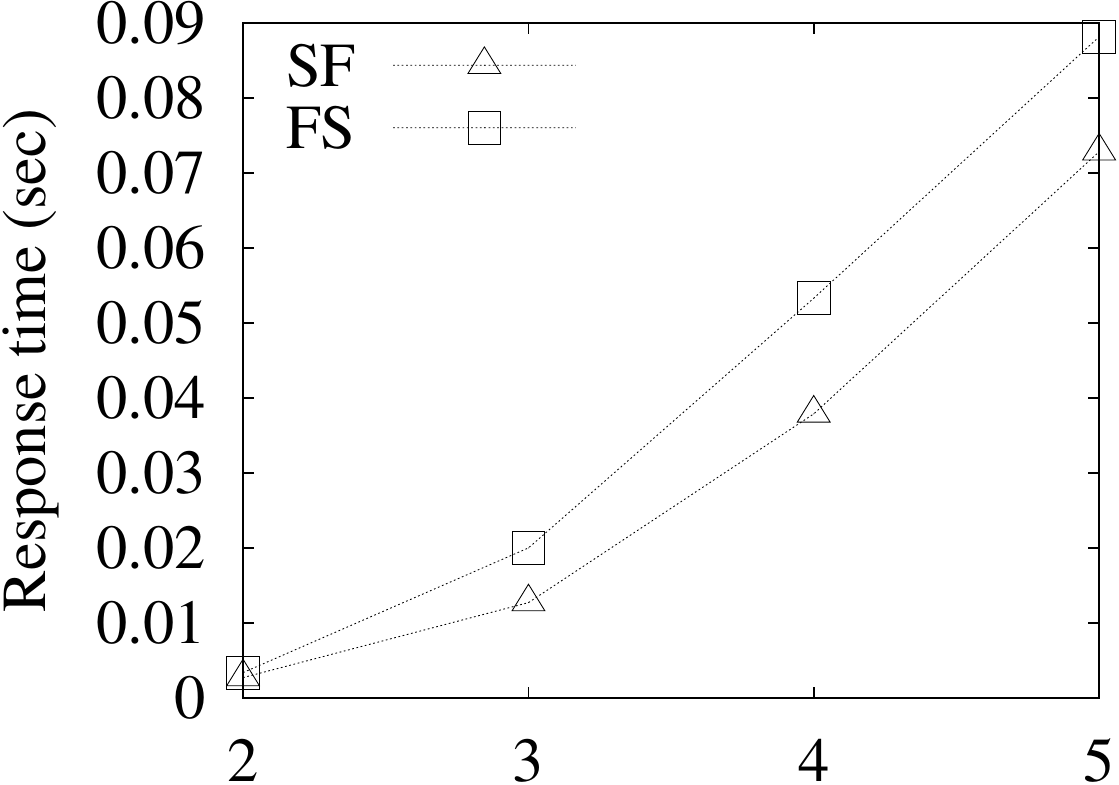}
&\hspace{-0.2cm}\includegraphics[width=0.245\linewidth]{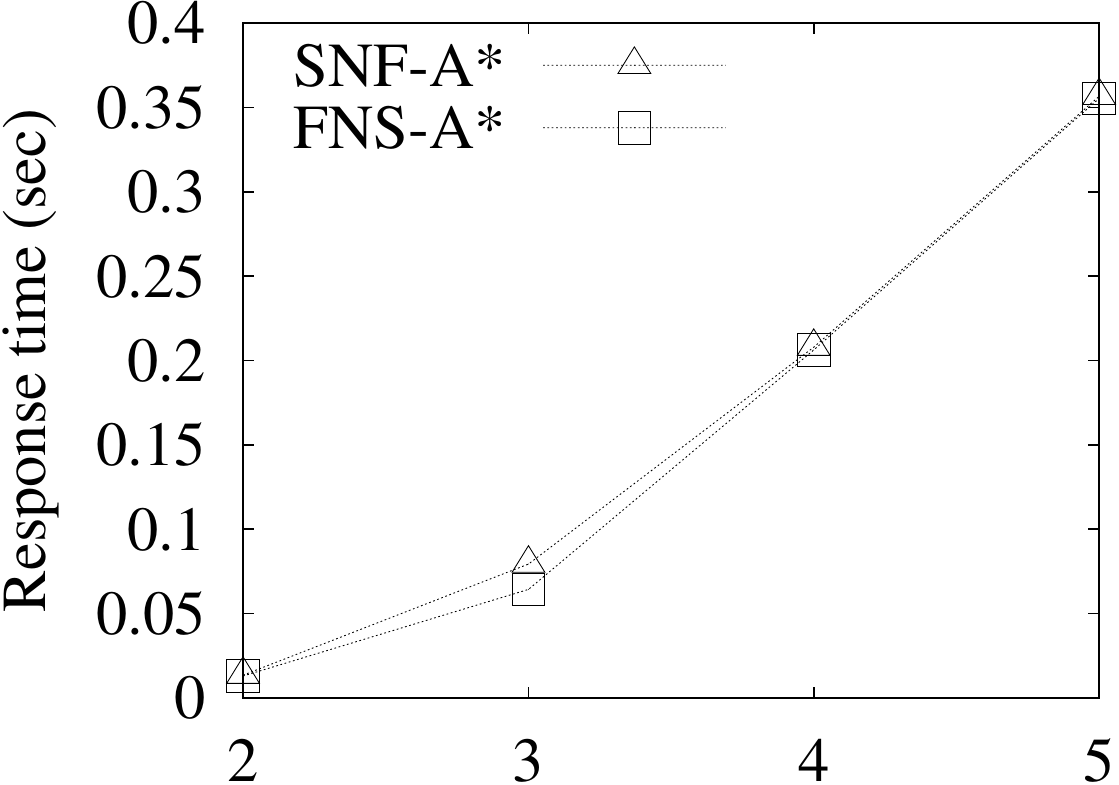}
&\hspace{-0.2cm}\includegraphics[width=0.245\linewidth]{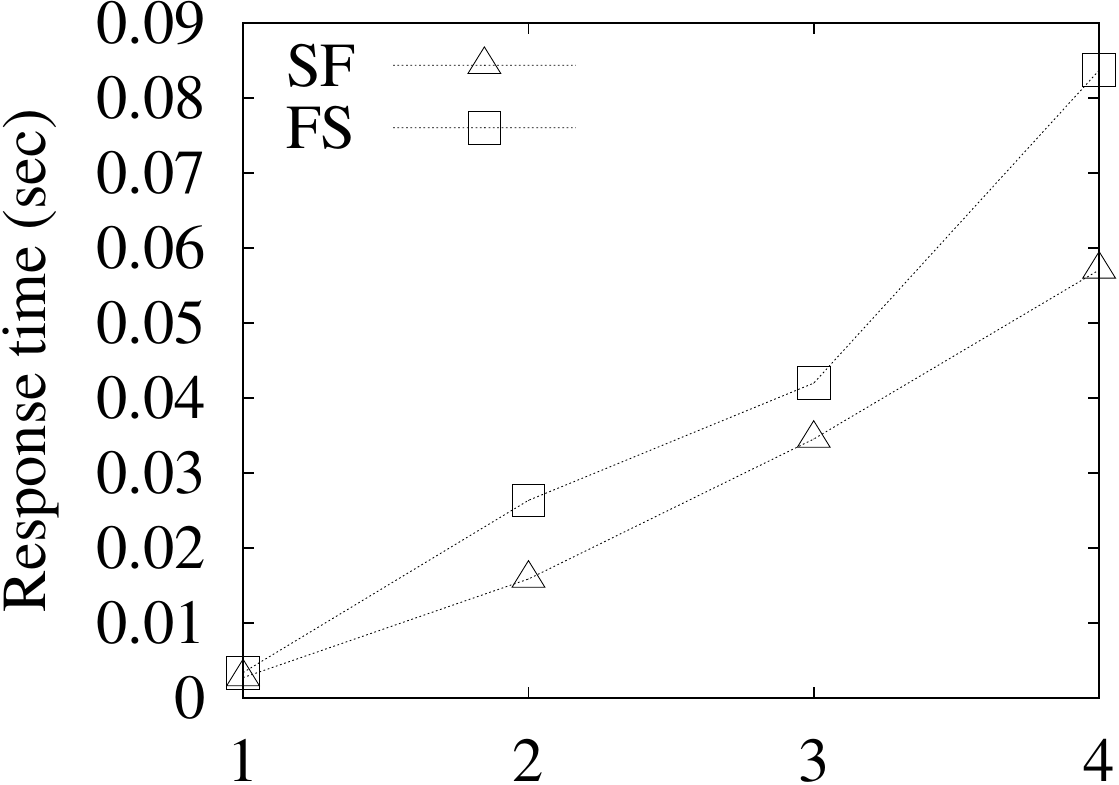}
&\hspace{-0.2cm}\includegraphics[width=0.245\linewidth]{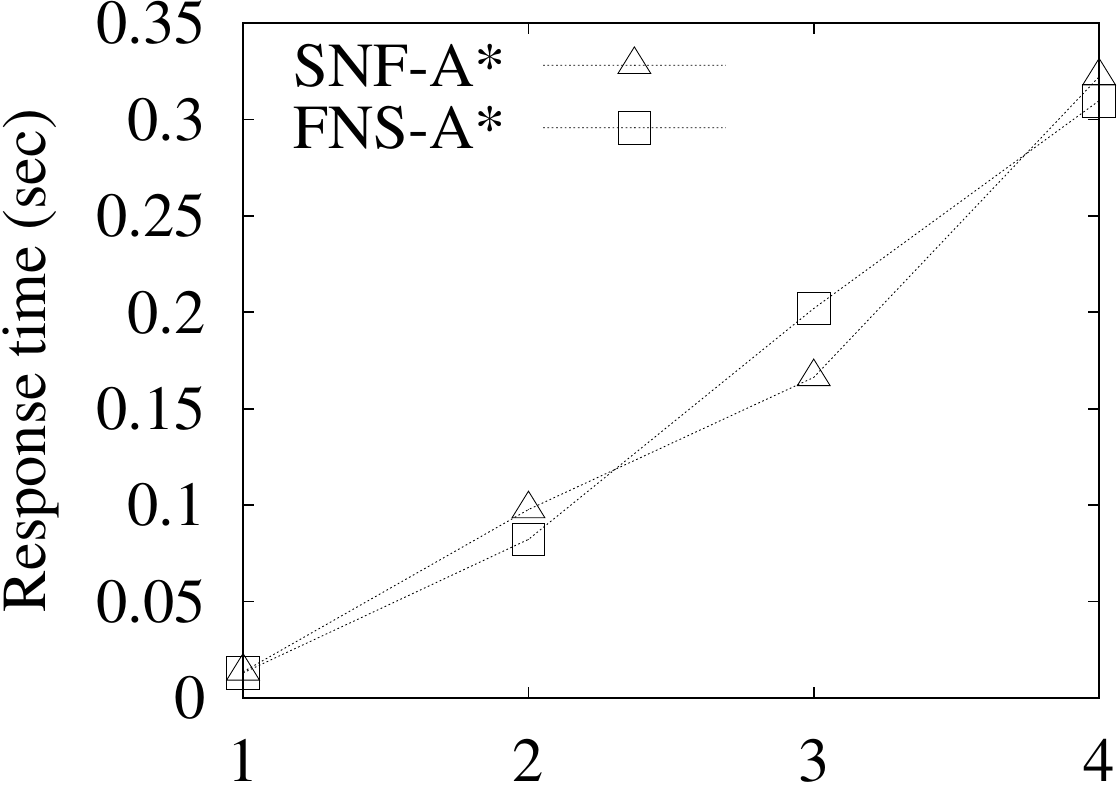}\\
%{\scriptsize Number of neighborhoods} &{\scriptsize Number of neighborhoods} &{\scriptsize Number of neighborhoods} &{\scriptsize Number of neighborhoods}\\
%$|N|$ &$|N|$ &$|N|$ &$|N|$\\
$\tau$ &$\tau$ &$\tau$ &$\tau$\\
(a) grid-based
&(b) grid-based
&(c) ring-based
&(d) ring-based\\
\end{tabular}
\vspace{-10pt}
\caption{Synthetic road networks: scalability tests for $\epsilon = 0.1$.}
\vspace{-10pt}
\label{fig:scalability}
\end{center}
\end{figure*}
In the last set of experiments we study the scalability of the best methods identified in the previous section, i.e., FS, SF, FNS-A$^*$ and SNF-A$^*$. For this purpose, we generate synthetic road networks varying the degree of the topology $\tau$\eat{number $|N|$ of neighborhoods included}, and thus, the size of the road network.
%Due to its construction, a grid-based backbone network takes $|N|$ values $\{1,4,9,16\}$, while a ring-based backbone $\{1,5,9,13\}$. 
Particularly, for a grid-based backbone network $\tau$ takes values inside $\{2,3,4,5\}$, while for a ring-based backbone inside $\{1,2,3,4\}$. 
Figure~\ref{fig:scalability} reports on the scalability tests. As expected, the response time of all methods increases when the degree of the topology\eat{the number of the neighborhoods, and thus the size of the road network,} increases. Even for the expensive fastest near-simplest and the simplest near-fastest route problems, our methods always identify the answer in less than half a second for $\epsilon = 0.1$. Although we do not include figures for other values of $\epsilon$, our experiments show that this holds for every other combination of $\tau$\eat{$|N|$} and $\epsilon$.
%!TEX root = main.tex

\section{Conclusion}
\label{sec:concl}

This paper dealt with finding routes that are as simple and as fast as
possible. In particular, it studied the fastest simplest, simplest fastest,
fastest near-simplest, and simplest near-fastest problems, and introduced
solutions to thems. The proposed algorithms are shown to be efficient and
practical in both real and synthetic datasets.

\stitle{Acknowledgments.}
This research was partially supported by the German Research Foundation (DFG) through the Research Training Group METRIK, grant no.\! GRK 1324, and the European Commission through the project ``SimpleFleet'', grant no.\! FP7-ICT-2011-SME-DCL-296423.

\bibliographystyle{abbrv}
\bibliography{biblio}

\end{document}